\documentclass[conference]{IEEEtran}

\IEEEoverridecommandlockouts  
\makeatletter
\def\@IEEEpubidpullup{} 
\def\@IEEEpubid#1{}     
\makeatother

\ifCLASSINFOpdf
\else
\fi

\usepackage{hyperref}
\usepackage[pdftex]{graphicx}
\usepackage[table,xcdraw]{xcolor}
\usepackage{enumitem}
\usepackage{listings}
\usepackage{array}
\usepackage{amsmath}
\usepackage{stfloats} 
\usepackage{tabularx} 
\usepackage{booktabs} 

\newcommand{\qt}[1]{\textit{#1}}

\newlist{tableitemize}{itemize}{1}
\setlist[tableitemize,1]{noitemsep, topsep=0pt, left=0pt, label=--}

\usepackage[switch]{lineno}

\newcommand{\revise}[1]{\textcolor{black}{#1}}

\newcommand{\datasetshort}{SENSE-7}
\newcommand{\datasetlong}{SENSE-7 (Subjective Empathy in Natural Sustained Exchanges; 7 dimensions)}

\newif\ifanonymous
\anonymousfalse  

\newcommand{\IfAnon}[2]{\ifanonymous #1\else #2\fi} 

\newcommand{\AnonText}[1]{\IfAnon{#1}{}}               
\newcommand{\RealText}[1]{\IfAnon{}{#1}}               
\newcommand{\SwapText}[2]{\IfAnon{#1}{#2}}             

\newcommand{\selfcite}[2]{\IfAnon{#1}{#2}}

\newcommand{\blindurl}[1]{\IfAnon{\texttt{[link hidden for review]}}{\url{#1}}}
\newcommand{\blindemail}[1]{\IfAnon{\texttt{[email hidden]}}{\texttt{#1}}}

\newif\ifieee
\ieeefalse   

\usepackage{pdfpages}

\begin{document}

\ifanonymous
\linenumbers
\else
\fi

\title{\revise{\datasetshort: Taxonomy and Dataset for} Measuring User Perceptions of Empathy in Sustained Human-AI Conversations}

\ifanonymous
\author{Anonymous Authors}
\else
\author{%
  Jina Suh\IEEEauthorrefmark{1},
  Lindy Le\IEEEauthorrefmark{2},
  Erfan Shayegani\IEEEauthorrefmark{3},
  Gonzalo Ramos\IEEEauthorrefmark{1}%
  \\  
  Judith Amores\IEEEauthorrefmark{1},
  Desmond C.~Ong\IEEEauthorrefmark{4},
  Mary Czerwinski\IEEEauthorrefmark{1}\IEEEauthorrefmark{5},
  Javier Hernandez\IEEEauthorrefmark{1}%
  \thanks{\IEEEauthorrefmark{1}Microsoft Research. Corresponding author: \texttt{jinsuh@microsoft.com}, \texttt{gonzo.ramos@gmail.com}, \texttt{judithamores@microsoft.com}, \texttt{javierh@microsoft.com}}%
  \thanks{\IEEEauthorrefmark{2}University of Michigan. \texttt{lindyle@umich.edu}}%
  \thanks{\IEEEauthorrefmark{3}University of California, Riverside. \texttt{sshay004@ucr.edu}}%
  \thanks{\IEEEauthorrefmark{4}University of Texas at Austin. \texttt{desmond.ong@utexas.edu}}%
  \thanks{\IEEEauthorrefmark{5}University of Washington. \texttt{marycz1031@gmail.com}}%
}
\fi

\markboth{Journal of \LaTeX\ Class Files,~Vol.~14, No.~8, August~2015}%
{Shell \MakeLowercase{\textit{et al.}}: Bare Demo of IEEEtran.cls for IEEE Journals}



\maketitle

\begin{abstract}
Empathy is increasingly recognized as a key factor in human–AI communication, yet conventional approaches to ``digital empathy'' often focus on simulating internal, human-like emotional states while overlooking the inherently subjective, contextual, and relational facets of empathy as perceived by users. In this work, we propose a human-centered taxonomy that emphasizes observable empathic behaviors \revise{and introduce a new dataset, \datasetshort, of  real-world conversations between information workers and Large Language Models (LLMs), which includes} per-turn empathy annotations directly from the users, \revise{along with} user characteristics, and contextual details\revise{, offering a more user-grounded representation of empathy.  Analysis of 695 conversations from 109 participants reveals that} empathy judgments are highly individualized, context-sensitive, and vulnerable to disruption when conversational continuity fails or user expectations go unmet. \revise{To promote further research, we provide a subset of \revise{672} anonymized conversation and provide exploratory classification analysis, showing that an LLM-based classifier can recognize 5 levels of empathy with an encouraging average Spearman \(\rho=0.369\) and Accuracy~\(=0.487\) over this set.} Overall, our findings underscore the need for AI designs that dynamically tailor empathic behaviors to user contexts and goals, offering a roadmap for future research and practical development of socially attuned, human-centered artificial agents.
\end{abstract}


%
\IEEEpeerreviewmaketitle

\section{Introduction}

Empathy is broadly conceptualized as the ability to recognize, understand, and share one another's emotions and perspectives~\cite{davis1983measuring, cuff2016empathy}.
Empathy is crucial to human communication and relationships, fostering trust, rapport, prosocial behaviors, and mutual understanding~\cite{ickes1993empathic, decety2011social, wondra2015appraisal}.
Motivated by the hypothesis that empathetic systems could lead to user engagement and satisfaction, researchers in human-computer interaction (HCI), human-robot interactions (HRI), and affective computing (AC) have explored how to embed empathic capabilities into artificial intelligence (AI) or to endow computing systems with ``artificial empathy''~\cite{picard2000affective}.
Indeed, recently developed Large Language Models (LLMs) can now produce empathic responses that people rate as more empathic than those written by humans~\cite{ayers2023comparing, rubin2025value, li2024skill, lee2024large, ovsyannikova2025third, wenger2025ai, yin2024ai}. 
There is a growing area of computing research around evaluating and improving these models' inherent ``empathy''~\cite{wang2023emotional,sorin2024large,lee2024large,welivita2024large, borg2024required, 9970384, 10899840}. 
At this juncture, re-examining empathy conceptualization and evaluation in human-AI interaction becomes essential.

Early empathic AI efforts drew from psychological theories~\cite{preston2002empathy} to simulate human emotional and cognitive processes, aiming \revise{to generate} cues to make agents ``as human-like as possible''~\cite{scotti2021modular}.
Prior studies modeled internal states and employed machine learning to convey empathic cues through facial expressions, voice prosody, linguistic strategies, and even haptic touch~\cite{paiva2017empathy, 5539766}, highlighting user benefits like comfort, support, or companionship~\cite{leite2013influence,tapus2007emulating,castellano2013towards,sharma2021towards,Chawla2021TowardsEA, Cuylenburg2021EmotionGA,vaidyam2019chatbots, bickmore2010response, 10316625}.
However, such efforts frequently overlooked the inherently subjective and contextual nature of empathy perception and accuracy~\cite{zaki2012neuroscience,ickes1993empathic}, raising feasibility and ethical concerns about objectively rating empathy from third-party perspectives~\cite{perry2023ai,montemayor2022principle,concannon2024measuring, 10388150}.

A growing body of work argues that empathy in human-AI interaction should instead be viewed as a dynamic, interactional construct centered on user experiences rather than AI's internal states~\cite{concannon2023interactional, borg2024required}.
This implies that the focus on empathy should be about how it is demonstrated through the AI's behaviors and communication strategies, rather than on any inherent features of the AI, like its ability to ``feel'' compassion.
Empathic behaviors can serve multiple functions, such as timely acknowledgments, context-sensitive clarifications, supportive emotional feedback, or proactive assistance, where each strategy might be employed at various points in a naturalistic interaction.
This shift in perspective also foregrounds and necessitates the user's subjective perception of empathy~\cite{inzlicht2024praise, hadjiandreou2025llmpathy}, as each strategy must be shaped in a way that resonates with the user's evolving goals, expectations, emotional states, and social contexts during a prolonged interaction~\cite{harrelson2020intention,gencc2024situating}.

Our work addresses gaps highlighted by this shifting perspective through a human-centered framework that emphasizes tailoring AI empathic behaviors to individual contexts and subjective perceptions.
\revise{Our primary contributions are twofold: (1)} a multidimensional \revise{taxonomy} of perceived empathic behaviors, \revise{which expands} traditional cognitive and affective empathy dimensions to include instrumental and exploratory functions, reframed as interactional and observable behaviors in AI agents\revise{; and (2) a new} human-centered digital empathy dataset, \revise{\datasetlong}, of English-based conversational interactions with general-purpose Large Language Models (LLMs). 
The dataset consists of subjective accounts of empathy from \revise{695 real-world} independent conversational interactions \revise{from} 109 information workers, including a range of individual characteristics, contextual and emotional states prior to and following interactions, per-conversation and per-response assessments of perceived empathy, and user preferences regarding different empathic dimensions.
\revise{As a secondary contribution, we present an exploratory classification analysis to demonstrate the utility of this taxonomy and dataset over a subset of 672 fully anonymized conversations\footnote{\SwapText{Instructions for obtaining a subset of the dataset scrubbed of personally identifiable information and confidential information can be found at [REDACTED].}{To inquire about the dataset or how to obtain it, please reach out to \href{mailto:sense7data@microsoft.com}{sense7data@microsoft.com}}}.}

\revise{From the evaluation of the data, we} found that both individual characteristics and conversational context significantly influence perceptions of empathy in AI interactions. Participants with higher daily AI use interest and higher trait empathy tended to select personal topics.
Participants with higher cognitive reappraisal scores rated their desired empathy level lower when dealing with personal or work issues compared to those with lower reappraisal scores. 
Analysis revealed that even a single ``poor'' turn in a conversation substantially diminished overall empathy ratings and correlated with lower engagement outcomes, despite similar conversation lengths across models.
By operationalizing \revise{perceived} empathy across \revise{seven} dimensions, we uncovered that users value nuanced distinctions, particularly favoring cognitive understanding and tailored responses. 
Participants' explanations highlighted  implicit expectations beyond conversation context, \revise{highlighting the} challenges \revise{of predicting empathy perceptions from} conversation alone. 
\revise{In this context, our exploratory classification experiments show that incorporating} conversational context \revise{via} adaptive-shot prompting strategy greatly enhanced the accuracy of detecting perceived empathy, highlighting the critical role of context in evaluating digital empathy.
\revise{Together, these} findings underscore the complexity and subjectivity of empathy perception in human-AI interaction. 
Our work contributes \revise{a tangible toolset} to the growing body of literature advocating for a human-centered paradigm for conceptualizing digital empathy, ultimately paving the way for AI agents that can dynamically adapt their empathic demonstrations to better meet the diverse needs of individual users.

\section{Related Work}\label{sec:related}

Empathy has traditionally been studied in psychology~\cite{cuff2016empathy} and communication research, defined as the ability to recognize, understand, and share another's emotional and mental states.
\revise{There are three general facets to empathy~\cite{zaki2012neuroscience}.
Two commonly distinguished facets are} affective (e.g.,~emotional resonance) and cognitive (e.g.,~perspective-taking) processes~\cite{davis1983measuring, decety2004functional}\revise{, with a third facet, prosocial motivation, recently gaining more attention~\cite{zaki2012neuroscience}.
The Affective component is largely concerned with feeling or resonating with someone's emotion~\cite{decety2004functional}, the Cognitive component is concerned with understanding someone's perspective or mental state~\cite{batson2009these, morgante2024possible}, and the Motivational component~\cite{eklund2021toward,zaki2014empathy} drives action or help~\cite{depow2021experience, guthridge2021taxonomy}.
Beyond these broad facets}, earlier works on human empathy attempt to capture fine-grained dimensions of empathy, such as comprehension~\cite{spreng2009toronto}, resonance or attunement~\cite{vachon2016fixing, decker2014development}, simulation~\cite{davis1983measuring, spreng2009toronto, reniers2011qcae}, concern~\cite{decker2014development}, emotional contagion~\cite{davis1983measuring,reniers2011qcae}, emotional sensitivity~\cite{spreng2009toronto}, physiological response~\cite{spreng2009toronto}, \revise{compassion~\cite{depow2021experience}, self-awareness~\cite{gerdes2010conceptualising}, fantasy~\cite{davis1983measuring}, emotional congruency~\cite{cuff2016empathy}, } or altruism~\cite{spreng2009toronto}. In addition, studies of human empathy tend to emphasize empathy's critical role in human relationships for building trust, rapport, prosocial behavior, and mutual understanding~\cite{ickes1993empathic, decety2011social, wondra2015appraisal}.

Motivated by these psychological studies, early affective computing and human-computer interaction research sought to imbue artificial systems with digital empathy-like qualities. 
Pioneering research treated empathy as something that could be simulated or added to AI systems by modeling human-like internal emotional states and expressing them through nonverbal and verbal cues~\cite{picard2000affective}. 
Social robots and embodied conversational agents attempted to convey empathy via facial expressions, gestures, voice prosody, haptic touch, or affective language to improve user comfort, trust, engagement, and prosocial action~\cite{breazeal2003toward, paiva2021empathy,brave2005computers,liu2024illusion,10316625,5539766}. 
For instance, systems designed to detect user emotions and respond with congruent affective displays have demonstrated several benefits. These empathic responses can enhance user experience, maintain social and emotional bonds with users across various domains including mental health support, education, and companionship~\cite{leite2013influence,tapus2007emulating,castellano2013towards,sharma2021towards,Chawla2021TowardsEA, Cuylenburg2021EmotionGA,vaidyam2019chatbots, bickmore2010response}.

Contemporary research on digital empathy centers on how empathy can be conceptualized, operationalized, and evaluated in human-machine communication~\cite{10388150, 9970384}. 
Although digital empathy is inspired by insights in human empathy, it is important to note that empirical work on human empathy often treats empathy as a stable internal state that can be measured through self-assessment scales given to the empathizer.
Few methods try to measure empathy through observation (e.g., by the recipient or by a third-party rater), especially in human-human interaction contexts where empathy is deemed more important, such as provider-patient interactions~\cite{decker2014development}.
Most empathy scales (given to the empathizer) use first-person statements such as ``I sometimes find it difficult to see things from the other guy's point of view~\cite{davis1983measuring}'' or ``It upsets me to see someone being treated disrespectfully~\cite{spreng2009toronto}.''
Such self-report methods are difficult to apply directly to AI because we cannot meaningfully interview or assess a system's \revise{``trait} empathy\revise{"}. 
\revise{Moreover, people also} vary widely in their empathic capacities and accuracies. Even within a single person, empathy fluctuates based on context\revise{, motivation,} or interpersonal dynamics~\cite{zaki2012neuroscience, zaki2014empathy, ickes1993empathic}. Therefore, some argue that directly applying human psychological trait models to AI and replicating human internal states in AI may be problematic. 

Individual differences (e.g.,~personality traits, cultural background, prior experiences) and contextual factors (e.g.,~the nature and goals of an interaction, emotional states) crucially shape how empathy is perceived and enacted~\cite{zaki2012neuroscience,davis1983measuring,eisenberg1983sex}.
In human-agent interactions, the relational and social role of empathy is more relevant than its manifestation as psychological traits for developing rich insights about users and helping address their needs, and this role of artificial empathy is highly contextual~\cite{concannon2023interactional}.
In addition, focusing on internal states can lead to anthropomorphizing AI systems, imputing them with the capacity to feel in human-like ways, which raises ethical and conceptual concerns~\cite{guzman2016making}.
Instead, scholars emphasize that empathy should be viewed as an interactional or relational phenomenon, realized through observable, context-sensitive behaviors rather than presumed internal mental processes~\cite{guzman2016making,concannon2023interactional, borg2024required}. 

In recent efforts to develop conversational AI agents with empathy, scholars have \revise{adapted} empathy dimensions from psychology \revise{and applied them in} third-person annotation \revise{schemes, where crowdworkers or experts label the presence of empathic cues in dialogues} as a scalable way to build \revise{training data}~\cite{rashkin2018towards, concannon2024measuring, sharma2021towards, sharma2020computational, xu2024multi, ziems2024can}.
\revise{This line of work has also motivated the development of} numerous taxonomies to characterize empathic responses or intents in human-AI dialogues. 
For instance, Welivita and Pu~\cite{welivita2020taxonomy} propose a large-scale taxonomy that highlights different empathy dimensions (e.g.,~emotional support, cognitive validation), while Svikhnushina et al.~\cite{svikhnushina2022taxonomy} demonstrate how questions that express interest or concern about a user's emotional state can be pivotal for perceived empathy.
\revise{Sharma et al.~\cite{sharma2020computational} operationalizes empathy into emotional reaction, interpretation, and exploration.
Increasingly, LLMs that power conversational agents are considered not only as the subject of evaluation but also as the evaluator itself~\cite{ziems2024can}.
For example, Kumar et al.~\cite{kumar2025large} evaluates the reliability of LLMs as judges for empathic communication using expert and crowd annotations consisting of these datasets.
While these approaches have advanced the field,} third-person labeling can neglect the subjective user experience (so-called second-person or participant viewpoint).
Directly obtaining feedback from the actual interactant on whether the agent was empathetic in that moment is the most ideal, but such methods pose practical challenges in data collection and experiment design~\cite{concannon2024measuring}.

\revise{A parallel stream of work has produced multi-turn and conversation-level empathy datasets since empathy in conversation often unfolds over multiple turns, where a user's emotional state and the agent's responses evolve dynamically~\cite{concannon2023interactional}. 
\revise{Each work leverages unique annotation schemes that emphasize different aspects of empathy in conversations.}
\textit{EmpatheticDialogues}~\cite{rashkin2018towards} collects crowdworker ratings of individual utterances. \textit{EPITOME}~\cite{sharma2020computational} captures seeker-responder pair annotations with highlighted rationales, \textit{ESConv}~\cite{liu2021towards} labels support strategies and outcomes, Xu and Jiang~\cite{xu2024multi} capture intent-empathy mappings from both speaker and listern perspectives, and Shen et al.~\cite{shen2024empathy} gather empathy ratings at the narrative level toward full stories authored by humans or AI. 
These diverse approaches illustrate the variety of ways empathy has been operationalized, while also leaving open opportunities to examine how perceptions of empathy evolve across turns in ongoing conversations.
}

Our work builds on this shift in perspective towards a human-centered approach to digital empathy that prioritizes empathy's functional role within interactional contexts~\cite{zhu2024toward}. \revise{To capture this} more accurately, our dataset \revise{provides both per‑turn and per‑conversation empathy judgments within multi-turn dialogues directly from the interactant, including} contextually relevant signals, such as \revise{participants'} current mood, expectations, \revise{and trait characteristics. 
This design enables analysis of how empathy unfolds across conversations, how a single turn can shape subsequence perceptions, and how} the broader interactional context \revise{influences} empathic perception.

\section{AI Empathic Behavior Scale}

Instead of emulating human empathy as an abstract concept, our goal is to apply understandings of human empathy in a way that is grounded in a particular human-AI interaction, tailoring AI's empathic behaviors to individual users and contexts.  
To achieve this human-centered approach to digital empathy, we first reframe traditional theories of cognitive and affective empathy in psychology as interactional, behaviorally observable phenomena in AI agents. 
Second, in addition to observable human-AI behaviors, we capture external factors, such as an individual's empathic accuracy or context, because these unseen external factors comprise the backdrop of empathic human-AI interactions.
Here, we describe development process of our empathic behavior scale. In our study, we use this scale in combination with individual characteristics and contextual factors to understand how AI behaviors unfold and are perceived as empathic by humans. 
Our scale is designed to measure text-based LLM behaviors, and we leave the extension of this scale to other modalities to future work.

We followed several steps during our development of a multidimensional AI empathic behavior scale.
First, we gathered existing psychology and neuroscience literature that defined components in human empathy (e.g.,~\cite{guthridge2021taxonomy}, \cite{cuff2016empathy}, \cite{depow2021experience}) as well as scales that measured human empathy (e.g.,~\cite{vachon2016fixing}, \cite{davis1983measuring}, \cite{lietz2011empathy}).
We also reviewed literature that examined applications of empathy in AI agents or robots (e.g.,~\cite{tapus2007emulating}, \cite{yalcin2018computational}) and scales that directly or indirectly measured empathy (e.g.,~\cite{fitrianie2022artificial}, \cite{sharma2020computational}).
These earlier works each had similar but unique ways of defining empathy components. 
Based on a social cognitive neuroscience understanding of empathy, the Empathy Assessment Index~\cite{lietz2011empathy} introduced five important components for empathy: Affective Sharing, Self-Awareness, Perspective Taking, Emotion Regulation, and Empathic Action.
For empathy in everyday life, Depow et. al.~\cite{depow2021experience} described the three distinct components of empathy as Emotion Sharing, Perspective Taking, and Compassion.
Outlining a computational model of empathy, Yalcin and DiPaola~\cite{yalcin2018computational} defined Communication Competence, Emotion Regulation, and Cognitive Mechanisms as three main components of empathy, along with mechanisms (e.g.,~Emotion Recognition, Theory of Mind) and behaviors (e.g.,~Empathic Concern, Altruistic Helping) associated with each component.
\revise{Supplementary Tables S1 and S2 enumerate all scales and articles we reviewed and the concepts introduced by them.}

From these works, \revise{we extracted individual scale items, empathy concepts, and their descriptions.}
\revise{Once} we gathered all concepts mentioned\revise{, two authors manually} conducted a conceptual mapping exercise (i.e., affinity diagramming \revise{and clustering based on conceptual similarity), where we found many smaller dimensions (e.g., resonance, congruency, perspective-taking, altruism), as enumerated in Supplementary Tables S1 and S2. We further grouped them into} three high-level categories---Affective (emotions), Cognitive (thoughts), and Motivational (intentions to help)---\revise{ and extracted 7 unique dimensions as follows}. 

Rather than describing empathic agents in terms of emotions, thoughts, and motivations that are not \revise{``}traits\revise{"} in \revise{artificial agents}, we \revise{instead reframed three main} concepts in terms of definable components of computing systems: their inputs, functions applied to that input, and outputs. As technology users do not have access to the algorithms underlying AI agents, we \revise{instead capture} the observable components from the human perspective: the inputs and outputs. Thus, the Affective and Cognitive empathy components are about whether the emotional and mental signals from the human (i.e., input \revise{to the agent}) are accurately identified and interpreted by the agent and whether the response produced by the agent (i.e., output \revise{of the agent}) meets the human's expectations (Affective Understanding and Cognitive Understanding). Likewise, \revise{we reframed} the Motivation \revise{empathy} component away from \revise{possibly making judgments about agent} intentionality, and \revise{instead} broke \revise{it} down into three observable behaviors regarding assistance that showed up as predominant themes in the affinity diagram: the actual response (Response Appropriateness), supportive behavior (Prosocial Expression), and actions signaling attunement (Interest). 
\revise{We included two additional dimensions under Motivation empathy to capture behaviors that demonstrate sustained effort through memory (Relational Continuity) and relevance (Contextual Understanding) for the following reasons.}
An interaction between the human and the agent is often multi-turned, composed of an ongoing loop of these inputs and outputs. This dynamic and constantly evolving aspect of empathy~\cite{main2017interpersonal} is captured by the Relational Continuity dimension. Moreover, the interaction between the human and the agent is situated within a broader social environment; Contextual Understanding describes the specific setting within which empathy occurs~\cite{guthridge2021taxonomy}.

\begin{table*}[ht]
\centering
\caption{Dimensions of Digital Empathy and Example AI Behaviors.}
\renewcommand{\arraystretch}{1.3} 
\setlength{\tabcolsep}{8pt}
\begin{tabular}{p{0.12\textwidth} >{\raggedright\arraybackslash}p{0.4\textwidth} >{\raggedright\arraybackslash}p{0.4\textwidth}}
\toprule
\textbf{Dimension} & \textbf{Description} & \textbf{Examples} \\
\midrule
\rowcolor{gray!15}\multicolumn{3}{l}{\revise{\textbf{Affective}}} \\

\textbf{Affective \mbox{Understanding} \revise{(e.g., \cite{depow2021experience, cuff2016empathy})}} &
The agent demonstrates the ability to recognize and understand my emotions/feelings. &
\textit{``I can sense you're feeling anxious right now. It's understandable to feel this way.''} \newline\textit{``I hear the frustration in your voice—I'm sorry you're going through that.''}
\\ 
\midrule

\rowcolor{gray!15}\multicolumn{3}{l}{\revise{\textbf{Cognitive}}} \\
\textbf{Cognitive \mbox{Understanding} \revise{(e.g., \cite{hall2021laypeople,eklund2021toward})}} &
The agent demonstrates the ability to recognize and understand my perspective/point of view, including goals and intentions. &
\textit{``So your main goal is to change careers—let's consider your interests and experiences.''}
\newline \textit{``I see where you're coming from; you want to ensure everyone's on the same page.''}
\\ 
\midrule

\rowcolor{gray!15}\multicolumn{3}{l}{\revise{\textbf{Motivational}}} \\

\textbf{Response \mbox{Appropriateness} \revise{(e.g., \cite{decker2014development,beredo2022hybrid})}} &
The agent demonstrates the ability to appropriately respond and adapt to my experiences, including when to provide advice or solutions. &
\revise{Rather than offering solutions directly, AI offers options: \textit{``That sounds very challenging. Would you like help identifying strategies or just talk through...?''}
\newline\textit{``That sounds tricky. If you're open to it, I can offer a few different perspectives.''}}\\

\textbf{Prosocial \mbox{Expression} \revise{\mbox{(e.g., \cite{vachon2016fixing,hall2021laypeople})}}} &
The agent demonstrates a concern for and a desire to help me. &
\textit{``I want to do everything I can to support you through this.''}
\newline \textit{``I really care about your well-being and want you to feel better.''}
\\ 

\textbf{Interest \revise{\mbox{(e.g., \cite{main2017interpersonal,decker2014development})}}} &
The agent demonstrates curiosity and attention toward my experiences. &
\textit{``That's fascinating—tell me more about how you got interested in this topic.''}
\newline \textit{``I'm genuinely curious about what led you to this point in your journey.''}
\\ 

\textbf{Contextual Understanding \revise{(e.g., \cite{guthridge2021taxonomy,main2017interpersonal})}} &
The agent demonstrates the ability to consider my unique circumstances (e.g.,~goals, beliefs, history, personality, preferences) and contextualize my experiences within broader external factors (e.g.,~culture, politics, social barriers, etc.). &
\textit{``Given your cultural background and past experiences, I can see how this might affect your decision.''}
\newline \textit{``Because you've told me you value honesty and clarity, I'll explain this in detail.''}\\

\textbf{Relational \mbox{Continuity} \revise{\mbox{(e.g., \cite{krupnick2006role,decker2014development})}}} &
The agent demonstrates the ability to maintain and enrich the relationship by consistently recalling and weaving details and nuances from past interactions into present exchanges, cultivating a dynamic and personalized rapport over time. &
\textit{``Last time we spoke, you mentioned a big presentation—how did it go?''}
\newline \textit{``I remember you were worried about your project deadline. Did you manage to wrap it up?''}
\\

\bottomrule
\end{tabular}
\label{tab:types_of_empathy}
\end{table*}

\subsection{Affective Understanding}
Affective understanding refers to an AI agent's ability to accurately sense and understand the human's emotional state. 
This abilty aligns closely with affective empathy or emotional empathy \revise{in human psychology} as ``the ability to accurately recognize others' feelings and understand the meaning of these feelings''~\cite{kalisch1973empathy,depow2021experience, cuff2016empathy}. 
When people believe their emotions are recognized, it creates a feeling of validation and emotional safety~\cite{kawamichi2015perceiving}.
Affective understanding is also a dimension that is most commonly studied and modeled as representative of empathy in NLP~\cite{concannon2024measuring, sharma2021towards,xu2024multi}.

Unlike human affective empathy where there is an emphasis on internal mirroring of feelings (i.e., resonance~\cite{decety2004functional,zaki2012neuroscience}), \revise{in our taxonomy} it is irrelevant that the agent has an internal mechanism to resonate emotionally. But it is still important that the user perceives that the agent is aware of and sensitive to the \revise{user's} emotions\revise{, and this dimension focuses on whether the user perceives behavior showing that the agent correctly understands their feelings. We keep this focus on observable behavior through the rest of the dimensions, below.}

\subsection{Cognitive Understanding}
Cognitive understanding refers to the ability of the AI agent to comprehend the human's perspective, intentions, and mental states. 
In human psychology, this aligns closely with cognitive empathy--a concept grounded in perspective-taking or theory of mind~\cite{carre2013basic,hall2021laypeople,eklund2021toward}.
Cognitive empathy is described as the capacity to mentally put oneself in someone else's shoes to understand their point of view~\cite{davis1983measuring}.
In the human-AI context, this dimension refers to the agent's demonstration that it grasps the reasons behind the user's feelings, the content of their concerns, or situations through paraphrasing what has been shared or reflecting meaning.

\subsection{Response Appropriateness}
Response appropriateness involves the AI agent's capacity to generate adaptive and context-sensitive responses.
Aligning prior research on therapist empathy~\cite{decker2014development} and dialog systems research~\cite{beredo2022hybrid}, response appropriateness is not just about recognizing emotions or points of view but about matching that recognition with a response that is both tactful and supportive.
It requires the agent to decide when to listen, validate, offer advice, or take another form of action that aligns with the user's needs.
In embodied agents, this could involve the expression of nonverbal cues that are contextually appropriate. 

\subsection{Prosocial Expression}
Prosocial expression involves the motivational aspect of empathy, especially the AI agent's desire to help and care for the user.
In human psychology, this may be similar to motivational empathy~\cite{zaki2012neuroscience}, empathic concern~\cite{davis1983measuring,vachon2016fixing}, or caring~\cite{hall2021laypeople,eklund2021toward}.
Although AI agents may not have internal mechanisms to feel concerns or care for humans, this dimension concerns the user's perception about AI's outward display to do something (i.e., responses and actions) that seems supportive, caring, and helpful.
It captures the extent to which the AI's \revise{expressed} empathy is active and human-oriented. In embodied agents, this could involve nonverbal actions like comforting touches when a human is upset or offering tissues.

\subsection{Interest}
Interest reflects an AI agent's active engagement in the ongoing interaction and curiosity regarding the user's experiences.
Prior research has shown that empathy is a dynamic process that unfolds over time~\cite{main2017interpersonal}, and the act of active listening and learning more about the other person builds a stronger connection and bonding and improves perception of empathy and relational warmth~\cite{hall2021laypeople,decker2014development,kawamichi2015perceiving}. \revise{Displaying such interest in a conversational partner is also an important component of active listening~\cite{jones2011supportive}.} 
In human-AI contexts, this could involve AI agents asking follow-up questions to explore~\cite{sharma2020computational} or providing back-channel cues (e.g.,~head nods, brief responses like ``uh-huh''~\cite{jang2024minimal}) to demonstrate its engagement and to emulate active listening.
Interest behaviors, therefore, demonstrate that AI agents have been paying attention.

\subsection{Contextual Understanding}
Contextual understanding pertains to integrating the user's broader background, such as personal history, cultural context, and preferences, into the empathic process. 
Empathy goes beyond emotions and perspectives; it depends on circumstances that form the situation under discussion~\cite{guthridge2021taxonomy,hall2021laypeople} as well as the relationship between the parties involved~\cite{main2017interpersonal}.
Context sets the common ground between the human and AI, which is important for AI to appropriately interpret and make sense of signals. 
Context is often not visible in the ongoing human-AI interaction but must be explicitly accessed and modeled as external knowledge~\cite{ma2020survey}.

\subsection{Relational Continuity}
Relational continuity focuses on the \revise{temporal} aspect of empathy and refers to the sustained, consistent empathic connection across multiple interactions, building a relationship rather than a one-off exchange. 
In counseling settings, therapeutic alliance has been shown to be important for improved perception of empathy and health outcomes~\cite{krupnick2006role,decker2014development}.
This is because providers accumulate knowledge about a patient that enables them to demonstrate a pattern of empathic attunement and genuine care.
Similarly, in human-AI contexts, forming a working relationship can lead to better adherence to interventions, engagement, and user satisfaction~\cite{bickmore2010maintaining}.
This dimension underscores the importance of memory for the accumulation of knowledge and continuity of interactions over time and across multiple interactions~\cite{main2017interpersonal, ma2020survey}.\newline

As an initial attempt at developing this scale, we formulated a single statement for each dimension.
These 7 dimensions form the core of our AI Empathic Behavior Scale, which we leverage and validate in our study. 
Table~\ref{tab:types_of_empathy} presents these statements along with examples.

\section{Human-Centered Empathic Conversation Data Collection} 
In the previous section, we defined 7 components of digital empathy, focusing on the interactional nature of empathy and various behavioral forms it can take in AI agents. In addition to the multidimensional aspects of empathy, contexts, especially individual characteristics, are important to consider when evaluating empathic behaviors of AI agents. For example, previous research has shown that people with a high empathic trait tend to have better empathic accuracy (i.e., their ability to perceive others' emotions and empathic responses)~\cite{ickes1993empathic,zaki2008takes}. 
Individuals' regulatory processes (i.e., emotion regulation skills) shape how we perceive and respond to emotions in others, which are key parts of empathy and interpersonal dynamics~\cite{thompson2019empathy}.
Individuals' attitudes towards AI may also influence how they interact with the AI (e.g.,~disclose private information) and rate AI's responses~\cite{wang2025understanding,cramer2010effects}.

To \revise{evaluate our taxonomy and to} understand how interactions with various AI systems can be perceived differently along empathy dimensions by different individuals, we conducted a study with 151 participants who engaged with the study's AI agent and assessed their perception of empathic behaviors in their assigned AI agent. 
To see how different AI agents and their behaviors would impact individual perceptions, we \revise{randomly} assigned participants  \revise{in a single-blind design} to one of four different AI agents (built on top of GPT-3.5-turbo, GPT-4, or Llama2-70b; see Section \ref{study:agent}). 
In each conversation, we obtained first-person and per-turn labels of empathy that enable investigation into within- and between-individual differences in the multidimensional perception of empathic behaviors in conversational AI. 
In addition, to further encourage more naturalistic conversation where individual differences will be drawn out, we conducted the study in real-world settings where individuals naturally incorporated the AI into their everyday situations. 
In this section, we describe the study design and data collection.

\subsection{Study Procedure}
We sent recruitment emails to a random sample of employees of a large technology company for their interest in participating in our study and enrolled participants on a first-come-first-served basis. 
Our study consisted of three main stages. (1) At intake, each participant completed an intake survey that captured their demographics and psychometrics~(\autoref{study:intake}). (2) Within the study period that lasted approximately 4 weeks, participants were asked to engage in at least 5 distinct conversations with an assigned AI agent whenever they felt the need for AI assistance~(\autoref{study:conversation}). (3) At exit, each participant completed an exit survey to reflect on their overall study experience and perceptions of empathy~(\autoref{study:exit}).
Our study employs a repeated-measures design in capturing multiple conversations and empathic perceptions per participant. 
It also employs a between-subjects design where each participant was assigned to one of the four different AI agents.
Each participant that completed the intake survey, 5 conversations, and the exit survey was compensated with a \$50 USD gift card. For each additional conversation, we compensated \$10 USD for a total compensation of up to \$100 USD per participant. The study was approved by the \SwapText{[REDACTED]}{Microsoft Research} Institutional Review Board.

\subsubsection{Intake survey}\label{study:intake}
The intake survey captured participants' demographics, such as age, gender identity, job function, highest attained education level, and whether English was their first language. 
We also captured psychometrics that may influence how individuals may interact with and perceive the AI agent. 
First, we captured the individual's empathic trait using two validated scales: (1) Single Item Trait Empathy Scale (SITES)~\cite{konrath2018development} and (2) Toronto
Empathy Questionnaire (TEQ)~\cite{spreng2009toronto}. 
Second, we captured the individual's emotional regulation skills using the Emotion Regulation Questionnaire (ERQ), with two subcomponents (cognitive reappraisal and expressive suppression)~\cite{gross2003individual}.
Third, we captured the individual's attitude towards AI using a combination of scales that were adapted for our study: (1) AI Attitude Scale (AIAS-4)~\cite{grassini2023development} which were mostly positively framed, (2) top two positive and two negative items with the highest factor loadings from the General Attitudes towards Artificial Intelligence Scale (GAAIS)~\cite{schepman2020initial}, and (3) our own statements (``I trust AI" and ``I would disclose personal information to AI") to capture attitude towards disclosure to AI. All AI attitude statements were answered on a 5-point agreement scale (1: Strongly disagree, 5: Strongly agree). 

\subsubsection{AI assignment}
We used data from the intake survey to conduct block randomization in assigning AI agents. In other words, we randomly distributed eligible participants while maintaining balance across four AI agents. 
We used four variables for block randomization: gender, AI attitude, TEQ, and expressive suppression subcomponent of ERQ. 
For AI attitude, TEQ, and expressive suppression, we used the median value to divide participants into high or low groups. 
Chi-square test of independence and ANOVA test revealed no significant difference across AI assignments along gender, SITES, TEQ, both subcomponents of ERQ, and AI attitude. 
Participants were blind to the AI assignment, and they were required to use the same AI agent for all of their subsequent conversations.

\subsubsection{Conversation task}\label{study:conversation}
After intake, participants were provided with instructions detailing how to access and interact with the AI agent, how to complete pre-conversation and post-conversation surveys, and how to label AI agent responses. Each participant was asked to engage in 5 or more conversations with their assigned AI agents for approximately 5 to 10 minutes. For participants' convenience, we sent a daily reminder with a link to the conversation task accessible via a browser. 
The conversation task consisted of 4 \revise{stages}: (1) pre-conversation survey, (2) conversation interface, (3) per-turn labeling interface, and (4) post-conversation survey.

\paragraph{Pre-conversation survey}
When the participant entered the conversation task, we captured momentary information about the participant before engaging with the AI agent. 
We asked participants to choose the topic of assistance needed among 9 topics grouped into 3 topic categories: information tasks (``I need help thinking through a work assignment", ``I need help with writing", ``I need help learning a new skill", or ``I need help with a personal task"), personal issues (``I need help overcoming a personal issue", ``I need help navigating a distressing situation", or ``I need help navigating a social situation"), and work-related issues (``I need help thinking through a work issue" or ``I need help thinking through a career/self-improvement problem").

We instructed participants to engage in a variety of topics, and to promote topic diversity, we showed the count next to each topic. 
We then asked them to write the task description and rate its importance (1: Not at all important, 5: Extremely important). 
We presented the definitions of the 7 empathy dimensions and asked them to rate their desired empathy level for the task (1: minimal, 2: moderate, 3: high) with an optional field to capture any comments about their desired level of empathy.
Finally, we asked them to assess their current mood through a 10-item international Positive and Negative Affect Schedule (I-PANAS-SF)~\cite{thompson2007development}.

\paragraph{Conversation interface}
After submitting the pre-conversation survey, participants were shown a text-based conversational chat interface~(Supplementary Figure S1). Participants were responsible for initiating the conversation with the AI agent. 
The interface included a text input box, a button to send the text, a loading indicator, and a scrollable window for showing the conversation history. There were no other functionalities to keep the interface as simple as possible.
Participants could terminate the conversation by clicking on ``I'm done" to navigate to the next \revise{stage} of the task.

\begin{figure}[t!]
  \centering
  \includegraphics[width=.9\linewidth]{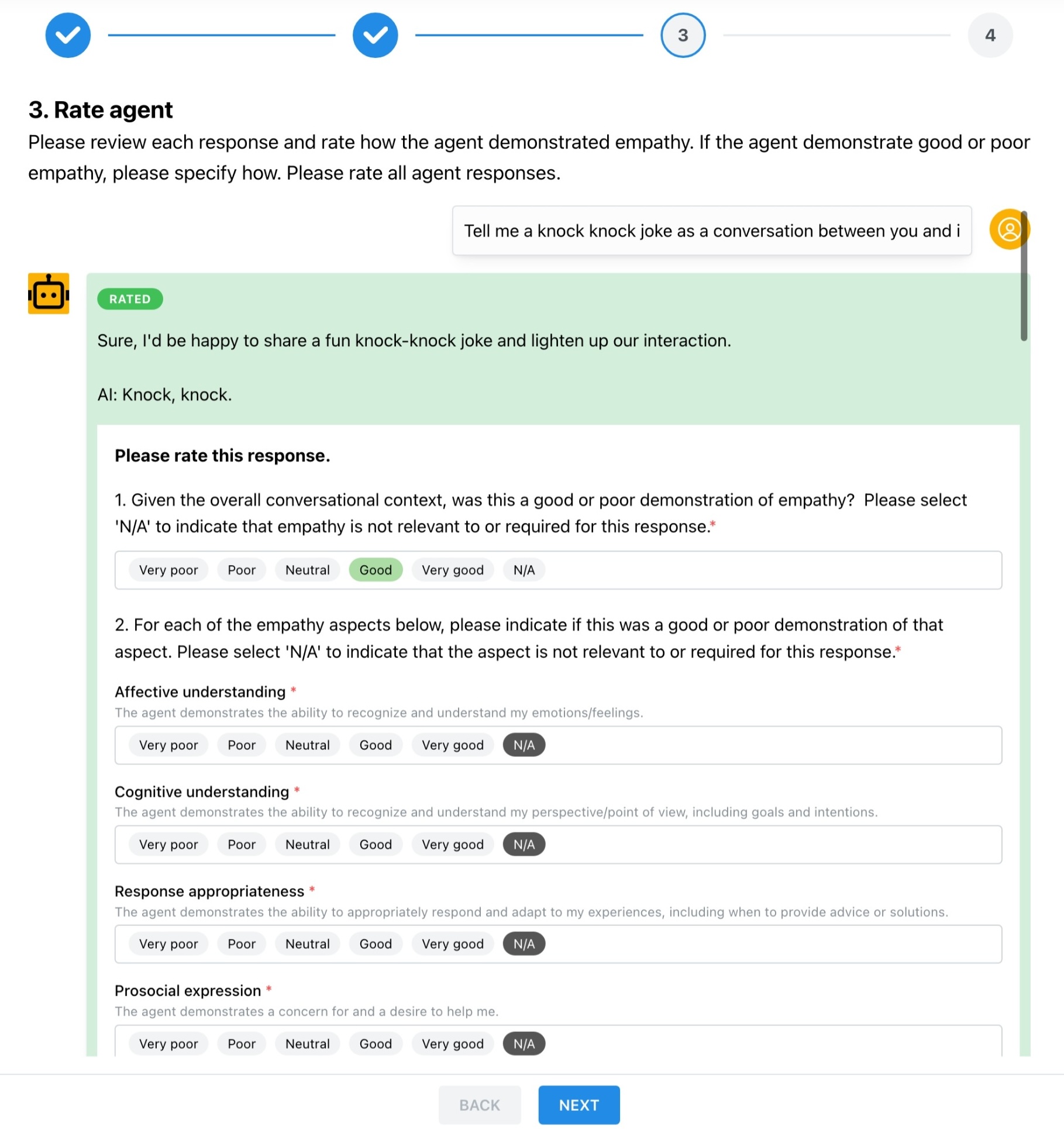}
  \caption{Per-turn labeling interface by the conversation participant}
  \label{fig:turn-labeling}
\end{figure}

\paragraph{Per-turn labeling interface}
After the conversation, we asked participants to label each of the AI responses for its overall demonstration of empathy, and along 7 dimensions, as shown in~\autoref{fig:turn-labeling}. Empathy labels ranged from `Very poor (1)' to `Very good (5)', with an option for `N/A' to indicate that empathy or a specific dimension of empathy is not relevant for that response.
Participants were required to label all AI responses before proceeding to the next \revise{stage}.

\paragraph{Post-conversation survey}
Lastly, we asked participants for their momentary mood again using I-PANAS-SF. 
We also asked participants to rate their conversational experience on whether they successfully completed the task, whether they were engaged or absorbed in the conversation, whether the interaction was positive, and whether they would use the agent again for the same task (1: Strongly disagree, 5: Strongly agree). 
We concluded the task by asking their overall perception of the agent's empathy along 7 dimensions and optional examples \revise{that justify} their ratings.

\subsubsection{Exit survey}\label{study:exit}
Questions in the study's exit survey were mostly open-ended.
We first asked participants to reflect on their experiences to comment on whether their desired level of empathy in AI agents has changed over the course of the study.
We asked participants to rank the 7 empathy components based on their importance overall\revise{, as well as their importance specifically in the context of} personalization\revise{--that is, which components were most essential to tailor to individual users}.
Finally, we asked participants to share their thoughts on designing AI agents to simulate or express human-like emotions, as well as designing AI agents to have human-like empathy.

\subsection{Conversational AI agent}\label{study:agent}
Our study used four different LLMs.
For the first instance (referred to as `GPT4'), we used Azure OpenAI's GPT-4 deployment (GPT-4-32k version 0613). 
For the second instance (referred to as `GPT4-empathy'), we used the same GPT-4 deployment with a system prompt, designed to encourage the agent to be more empathic, including reflective questions based on the 7 dimensions of empathy~(Supplementary Table 1).
For the third instance (referred to as `Llama2-70b'), we used Llama2-Chat-70B version hosted on Azure.
For the fourth instance (referred to as `IC'), we used GPT-3.5-turbo, where each turn was preceded with a prompt to extract the user's evolving intent and information to remember (e.g.,~memory) which was then provided as additional context to generate the best response.
All instances used similar settings, such as a temperature of 0.1 and top-$p$ of 0.9. 

\subsection{Study population}
From our call for participation, 247 people expressed interest and consented to participate in the study, and 197 participants completed the intake survey.
Of the 197 participants, 151 participants submitted at least one conversation, and 109 participants submitted at least five conversations and completed the exit survey.
From these 109 participants, we obtained 695 fully labeled conversations.
Of 109 participants, 55 identified as men, and 54 identified as women. 
33 participants (30.3\%) were aged 26–35 years, 31 (28.4\%) were 46–55 years, 24 (22.0\%) were 36–45 years, 10 (9.2\%) were 56–65 years, and 8 (7.3\%) were 18–25 years old, with 3 (2.8\%) preferring not to disclose their age.
46 participants (42.2\%) had a Bachelor's degree as their highest education level, 43 (39.4\%) had a Master's degree, 10 (9.2\%) had some college or vocational training, 5 (4.6\%) had doctoral or professional degrees, and 2 (1.8\%) had some postgraduate degree, with 3 (2.8\%) preferring not to disclose their education.
81 participants (71.3\%) indicated that English was their first language.
The three largest role categories were software development and engineering (42, 38.5\%), business and finance (27, 24.8\%), and product and project management (19, 17.4\%), with others including data analytics, research, design, administration, and support roles. 

Based on psychometrics data captured from the intake survey, participants self-assessed that they were highly empathetic ($\bar{x}$=4.37 out of 5, $\sigma$=0.66), although the average TEQ score ($\bar{x}$=41.0, $\sigma$=7.12) indicates lower empathy compared to the reported means in the original scale development ($\bar{x}$=46.25~\cite{spreng2009toronto}).
The distribution of ERQ scores confirms prior research~\cite{grassini2023development}, where we saw statistically significant gender differences for expressive suppression but not for cognitive reappraisal. 
In our study population, men suppressed emotions more than women ($\bar{x}_\text{man}$=3.85, $\bar{x}_\text{woman}$=3.25 out of 7, \emph{t}(107)=2.82, \emph{p}=0.006). The average cognitive reappraisal scores were 5.13 ($\sigma$=0.91) out of 7 for men and 5.23 ($\sigma$=0.94) for women.
The average AI attitude leaned towards positive ($\bar{x}$=3.78 out of 5, $\sigma$=0.47), with statistically significant gender differences ($\bar{x}_\text{man}$=3.96, $\bar{x}_\text{woman}$=3.61 out of 5, \emph{t}(107)=2.82, \emph{p}$\ll$0.001).
\revise{Potential over-representation of tech workers may have contributed to a skew toward positive AI attitudes.}

\subsection{Analysis}
\subsubsection{Quantitative}
For quantitative analysis, we employed various statistical methods using \texttt{Python} and its libraries (e.g.,~\texttt{pandas}, \texttt{scipy}, \texttt{statsmodels}, \texttt{pingouin}, \texttt{seaborn}).
When modeling for binary outcomes (e.g.,~participant chose a Personal Issue), we used a population-averaged Generalized Estimating Equations (GEE) model while accounting for the within-participant correlation by clustering on participant ID and leaving model assignment, individual characteristics, and contextual factors as fixed variables. 
We used linear mixed-effects models to understand the contribution of fixed variables (e.g.,~individual characteristics, contextual factors) on various outcome variables (e.g.,~desired empathy, perceived empathy), treating participant identifiers as random variables.
When reviewing differences in means across groups, we used t-tests for two groups and ANOVA tests for three or more groups. We used the Benjamini-Hochberg procedure for controlling the False Discovery Rate (FDR) and to correct for multiple comparisons. Where applicable, we conducted the Tukey Honest Significant Difference (HSD) procedure for pairwise comparisons.
We calculated consistency across variables using Cronbach's alpha, agreement between two variables using Cohen's Kappa, effect sizes using Cohen's alpha, and correlations using Pearson's r. 

\subsubsection{Qualitative}
To understand how each empathy dimension is demonstrated well or poorly, one author prioritized and qualitatively reviewed conversations with low or high empathy ratings \revise{in conjunction with the subset of participants who provided optional comments}. The second author reviewed and confirmed qualitative interpretations of the comments. 
\revise{Because comments were not required, our analysis does not aim for thematic saturation, but instead uses these comments as illustrative cases that clarify how participants interpreted chatbot behaviors and how these interpretations informed their ratings.}
Participant comments were crucial to validate our observations of the conversation and \revise{to uncover} misaligned expectations. 
For analyzing exit survey responses, the same author reviewed the open-ended responses using reflexive thematic analysis~\cite{merriam2002introduction}, specifically focusing on \revise{highlighting diverse rationales for} why certain empathy dimensions were perceived to be important\revise{, rather than claiming comprehensive coverage of all perspectives}. 

\subsubsection{Perceived Empathy Recognition} \label{subsec:analysis}

\revise{To evaluate the feasibility of automatically measuring perceived empathy at a conversation level, we explored the use of a simple LLM based classifier with the GPT 4o Azure OpenAI service (\textit{api\_version} 2024-10-21). The prompt included in Supplementary Table S4 was structured in three parts. First, a goal definition focused on analyzing the quality of the assistant responses in context. Second, an empathy description that incorporated the seven dimensions and their definitions in Table~\ref{tab:types_of_empathy}. Third, a five point rating guide as shown in Figure~\ref{fig:turn-labeling}. To promote deterministic behavior we set \textit{top\_p} to 1.0 and set \textit{temperature}, \textit{presence\_penalty}, and \textit{frequency\_penalty} to 0.0.}

\revise{To obtain the ground truth of perceived empathy, we first estimated the empathy level for each turn by averaging the seven turn level dimensions, which yielded one score per turn. We then computed the average of these scores to estimate the perceived empathy at a conversation level. The score range covered the full scale with minimum 1 and maximum 5. To ensure potential replicability, this part of the analysis was performed over a smaller subset of 672 conversations that could be fully anonymized. The remaining 23 conversations were deemed too sensitive and difficult to anonymize while preserving the semantic meaning of the conversation such as participants sharing confidential information or iteratively refining personal emails (see Supplementary Section 4 for further details on the anonymization process). The final set of anonymized conversations that are shared with this paper contained \(N=672\) conversations distributed as follows. Very Poor~\(=5\), Poor~\(=15\), Neutral~\(=48\), Good~\(=290\), and Very Good~\(=314\).}

\revise{We evaluated performance by computing Mean Absolute Error (MAE) and Spearman correlation using the continuous predictions. We then rounded predictions and labels to compute and report Accuracy, Macro Sensitivity, Macro Specificity, and Macro F1. To quantify model variability, predictions for all conversations were obtained ten times and results were reported as mean and standard deviation across runs. Part of the results include a regression scatter with a 95\% prediction interval that was computed using the standard linear regression prediction interval approach with a critical $t$ value, residual standard error, sample size, and a leverage term. In addition, we also provide the empirical cumulative distribution of the absolute level gap \(D=|\hat{y}-y|\) over \(\{0,1,2,3,4\}\). The full conversation context was passed to the model and no truncation was necessary. 
The goal of this analysis was not to propose a state of the art classifier, but to assess the feasibility of empathy classification using context sensitive prompting.}

\section{Results}

We organize our findings from this data into four sections. 
We first analyze how individual and contextual differences influence perception of empathy (without going into the conversation) (Section~\ref{results:quant-factors}).
We review the conversations qualitatively to understand how empathy dimensions show up (Section~\ref{results:qual-dimensions}).
We then analyze exit survey responses to understand the role of empathy dimensions in AI systems (Section~\ref{results:qual-role}). 
Finally, we provide some results in the context of perceived empathy recognition in conversations (Section~\ref{results:prediction}).

\subsection{What factors influence perceptions of empathy?}\label{results:quant-factors}

Our quantitative analysis of post-task perceived empathy examines pre-task factors (e.g.,~conversation topics, desired empathy levels, task importance), conversational factors (e.g.,~conversation length, turn-by-turn empathy rating), and combines these with static factors (e.g.,~individual characteristics, model types) to model the influence on post-task empathy ratings. 
We leave more in-depth analysis of the conversational dynamics and styles for Section~\ref{results:qual-dimensions}.

\subsubsection{Conversation topics and user expectations}
Before engaging in a conversation with the chatbot, we asked participants to categorize the conversation into one of three categories (information, personal, work) based on assistance needed. Of the 695 conversations, 43.7\% were information tasks, 30.4\% were personal issues, and 25.9\% were work-related issues.

From our GEE analysis of choosing personal issues, we found that, on average across participants, those with higher daily AI use interest ($\beta$~=~0.27, $p$~=~0.021, OR $\approx$ 1.31) and trait empathy ($\beta$~=~0.30, $p$~=~0.042, OR $\approx$ 1.35) were more likely to pick personal issues on average, whereas those perceiving organizational AI use as unethical ($\beta$~=~-0.19, $p$~=~.016, OR $\approx$ 0.82) or expressing greater willingness to disclose personal information ($\beta$~=~-0.13, $p$~=~.042, OR $\approx$ 0.87) were less likely to do so.

For work issues, we found that on average, the participants who were significantly more likely to choose a work issue were those in the 46--55 age group ($\beta$~=~0.68, $p$~=~.047, OR $\approx$ 1.96), those preferring not to disclose their age ($\beta$~=~1.23, $p$~=~.003, OR $\approx$ 3.41), those who perceive 
organizational AI use as unethical ($\beta$~=~0.49, $p$~$<$~.001, OR $\approx$ 1.64), those who plan to use AI ($\beta$~=~0.58, $p$~=~.002, OR $\approx$ 1.78), and those willing to disclose personal information ($\beta$~=~0.17, $p$~=~.047, OR $\approx$ 1.18). 
In contrast, the participants who were less likely to select a work issue on average were those presented with the `llama2-70b' model ($\beta$~=~-0.50, $p$~=~.002, OR $\approx$ 0.61), those who believe AI applications are beneficial ($\beta$~=~-0.27, $p$~=~.026, OR $\approx$ 0.76), those who trust AI ($\beta$~=~-0.28, $p$~=~.039, OR $\approx$ 0.75), and those higher in trait empathy ($\beta$~=~-0.58, $p$~$<$~.001, OR $\approx$ 0.56). \revise{We note that conversations with `llama2-70b' for work issues had the lowest average post-empathy rating, although only significantly lower compared to `GPT4-empathy' in pairwise comparisons. Linguistic differences across models~\cite{lee2024large} may have contributed to the selection and perception of empathy.}

For information tasks, we found that participants in the 26--35 age group ($\beta$~=~-0.54, $p$~=~.007, OR $\approx$ 0.58), those aged 46--55 ($\beta$~=~-0.55, $p$~=~.001, OR $\approx$ 0.58), and those who preferred not to report their age ($\beta$~=~-1.20, $p$~=~.002, OR $\approx$ 0.30) were less likely to select an information task, on average. Additionally, participants who believed that AI is positive for humanity ($\beta$~=~-0.20, $p$~=~.046, OR $\approx$ 0.82) or that organizations use AI unethically ($\beta$~=~-0.16, $p$~=~.049, OR $\approx$ 0.85) also showed lower odds of choosing an information task. 
These findings suggest that individual characteristics and model types do influence people's choice of conversational topics.

\begin{figure}[t!]
  \centering
  \includegraphics[width=\linewidth]{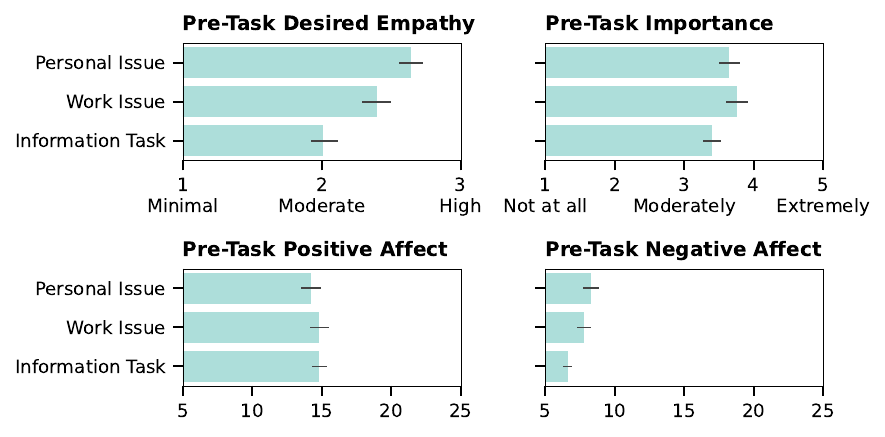}
  \caption{Desired empathy, importance, and mood per conversational topic\revise{. Affect scale ranges from 5 to 25, with 25 representing the max amount of that affect type.}}
  \label{fig:desired}
\end{figure}

We found statistically significant differences across the three conversational topic areas in terms of the desired empathy level, perceived importance of the conversation, and negative affect before the conversations (Figure~\ref{fig:desired}). 
When we looked at how individual characteristics and topic choice influenced the desired level of empathy, task importance, and mood for the conversation about to be had, we generally found that participants who included personal or work issues experienced higher emotional intensity, saw the conversation as more important, and sought more empathy.
Specifically, participants who were older (26--35: $\beta$=0.397, $p$=.003; 46--55: $\beta$=0.309, $p$=.024; 56--65: $\beta$=0.312, $p$=.048), those who chose a personal ($\beta$=0.679, $p$$<$.001) or work ($\beta$=0.408, $p$$<$.001) issue, and those believing AI applications are beneficial ($\beta$=0.153, $p$=.035) reported higher desired empathy.
Those tackling a personal ($\beta$=0.236, $p$=.004) or work ($\beta$=0.302, $p$$<$.001) 
issue perceived the task as more important, whereas higher trust in AI 
($\beta$=-0.243, $p$=.030) corresponded to lower importance.
Choosing a personal issue ($\beta$=-0.622, $p$=.026) was associated with lower positive affect.
Participants tackling personal ($\beta$=1.621, $p$$<$.001) or work ($\beta$=1.115, $p$$<$.001) issues showed heightened negative emotions.
Higher cognitive reappraisal was associated with higher positive affect ($\beta$=0.907, $p$=.041) and lower negative affect ($\beta$=-0.623, $p$=.011).

We then looked at what individual characteristics may moderate the relationship between the topic choice and the desired empathy level.
In addition to the main effect of topic choice, we found that participants with higher scores in cognitive reappraisal tend to desire slightly more empathy from AI systems across all types of tasks ($\beta$=0.107, SE=0.051, p=0.037).
Among interaction terms, we found a negative interaction between personal ($\beta$=-0.151, SE=0.067, p=0.024) or work issues ($\beta$=-0.160, SE=0.069, p=0.021) and cognitive reappraisal.
This suggests that participants with higher cognitive reappraisal scores rated their desired empathy level lower when dealing with personal or work issues compared to those with lower reappraisal scores.

\subsubsection{Conversation length and turn-by-turn empathy ratings}
Across 695 conversations, we find that participants had on average 10.0 turns ($\sigma$=6.2) with average agent response of 1368.7 characters per turn ($\sigma$=621.4).  
Although we found no statistical difference in the number of turns per conversation across models, we found average agent response length to be significantly different across models (f=151.32, p~$<<$~0.001), with `llama2-70b' having the largest character count (1826.7) and `IC' having the smallest (650.6). 
We found no significant effect of agent character length on empathy rating, both at turn-by-turn level and at the conversation level, suggesting that there are other factors (e.g.,~conversational style) that might influence the perception of empathy.

Of 6997 turns within 695 conversations, participants labeled 45.9\% of turns as applicable for empathy. Among the dimensions, Cognitive had the highest (43.0\%) and Relational had the lowest applicability (38.1\%). 
When we look at turn-by-turn empathy ratings for 2324 turns where all 7 empathy dimensions applied, the ratings on all the dimensions were internally consistent ($\alpha$= 0.961; 95\% CI [0.958, 0.963]).
However, there were subtle differences along the dimensions, which should not be ignored.
Of 2324 turns, 195 turns (8.4\%) had conflicting ratings among the 7 dimensions (i.e., ``poor" on one dimension and ``good" on another).
12.7\% of turns had at least one ``poor" rating (hereafter meaning Poor or Very Poor), and only 2.4\% had all 7 dimensions labeled as ``poor". 
The dimension with the most frequent `Very Poor' turns was Relational Continuity (2.2\%), and Prosocial had the least frequent `Very Poor' turns (1.1\%).
Average agreement between the overall empathy and individual dimensions was 0.617, with the highest agreement in Affective ($\kappa$=0.703) and the lowest agreement in the Relational dimension ($\kappa$=0.541).

We tagged each conversation as having a poor turn if there exists at least one ``poor" rating in any of the turns within the conversation. 
Between conversations with a poor turn and without, the mean differences in the number of turns ($p$~$<<$~0.001, without mean=9.5, with mean=11.7, f=17.03) and in the average user character length per turn ($p$=0.002, without mean=140.8, with mean=190.4, f=9.93) were statistically significant with small effect sizes ($d$=0.362 and 0.276, respectively), but there was no significant difference in the average agent character length per turn ($p$=0.812, without mean=1372.0, with mean=1359.0, f=0.06).

\subsubsection{Post-task perceptions}

\begin{figure}[t!]
  \centering
  \includegraphics[width=\linewidth]{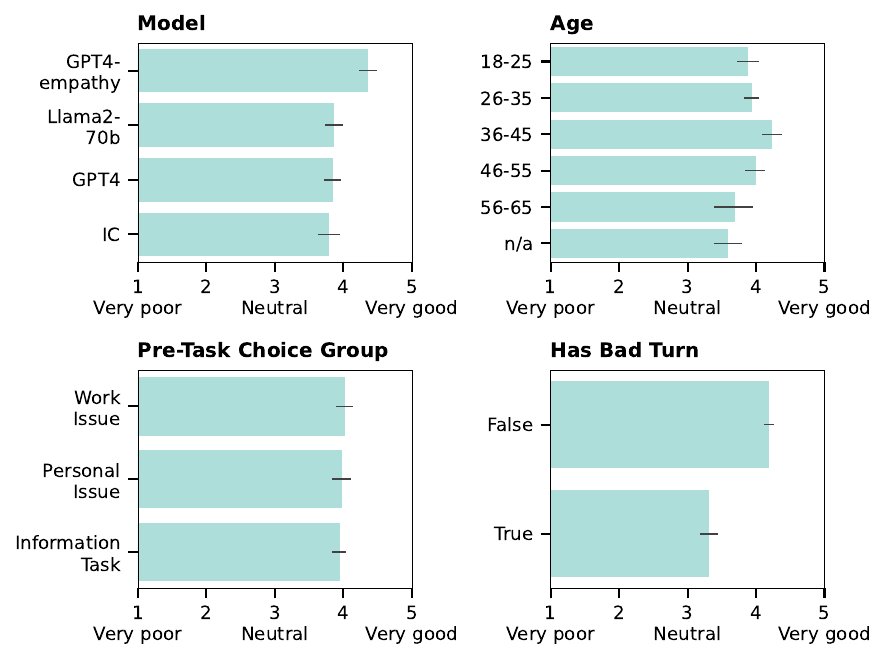}
  \caption{Post-task empathy rating across model, age, pre-task choice group, and having a bad turn}
  \label{fig:postempathy}
\end{figure}

When we examined post-task ratings along the 7 dimensions of empathy, we observed high consistency ($\alpha$~=~0.94; 95\% CI [0.933, 0.946]), indicating that these 7 items behave very coherently as a single scale and may point to one underlying construct. 
When we conduct pairwise comparisons, the Response dimension (mean=4.13, $\sigma$=0.96) had a significantly higher mean than Relational (mean=3.83, $\sigma$=1.12), Interest (mean=3.87, $\sigma$=1.04), and Contextual (mean=3.85, $\sigma$=1.09).
Because of the high consistency across dimensions, we compute post-task Overall rating by taking the average of the 7 ratings.

We modeled the Overall post-task empathy rating as a function of Model Type, individual characteristics, pre-task and conversational factors while treating participant ID as random effects (conditional R$^2$: 0.498, marginal $^2$: 0.294). 
We found that Model Type ($\beta$=[-0.35, -0.47] with `gpt4-empathy' as reference, all $p$~$<$~.05), having a poor turn ($\beta$=-0.657, $p$~$<<$~0.001), Trust AI ($\beta$~=~0.199, $p$~=~.029), 
and GAAIS: Interested Daily AI Use ($\beta$~=~-0.243, $p$~=~.014) were statistically significant predictors. 
Table~\ref{tab:lme_results_overall} presents coefficients for all predictors in the model. 
The effect size of having a ``poor" turn was large ($d$=1.142).
Through pairwise comparisons, we found `GPT4-empathy' to have a significantly higher rating compared to other models, and there were no statistically significant differences among the other three models.
We then examined how having a ``poor" turn might influence the overall empathy rating across Age or Gender groups. We found that, while being in the age group of 36-45 positively influenced the overall empathy ($\beta$=0.424, p=0.038), when this group was experienced a poor agent response, their empathy rating was negatively influenced ($\beta$=-0.590, $p$=0.028), suggesting that being in the 36–45 age group exacerbates the detrimental impact of having a poor turn on perceived empathy.
\revise{We did not find a significant effect of desired empathy rating on the overall perceived empathy. To investigate this further, we conducted a mediator analysis to understand the effect of desired empathy on the relationship between having a bad turn and the overall empathy rating and found no significant mediation effect.}
Figure~\ref{fig:postempathy} illustrates the average empathy rating per model, age, conversation topic, and having a poor turn.

When we compare post-task empathy ratings with four engagement outcomes, Overall empathy ratings were positively correlated with the task being successful (Pearson $r$=0.619), having an engaged experience ($r$=0.632), having a positive interaction ($r$=0.657), and wanting to use the agent again for the same task ($r$=0.602). 
Similarly to empathy, having a poor turn was a significant predictor of negative outcomes on all four engagement variables.
We also found age to be a significant moderator, such that, for individuals aged 26–55, having a poor turn substantially reduces their willingness to use the service again (by roughly 0.86 to 0.97 points).

\begin{table}[h!]
\centering
\caption{Mixed Linear Model Regression Results for Post-Task Overall Empathy Rating}
\label{tab:lme_results_overall}
  \begin{tabular}{lrr}
    \toprule
    \textbf{Parameter} & \textbf{Coefficient} & \textbf{P-value} \\
    \midrule
    Intercept & 4.151 & 0.000 \\
    \midrule
    \textbf{Participant-Based} & & \\ 
    \textit{Scenario Reference~=~GPT4-empathy} & &\\
    Scenario~=~GPT4 & -0.350 & 0.015 \\
    Scenario~=~IC    & -0.412 & 0.007 \\
    Scenario~=~Llama2-70b & -0.469 & 0.001 \\
    \textit{Gender Reference~=~Man} & &\\
    Gender~=~Woman           & 0.108  & 0.339 \\
    \textit{Age Reference~=~18–25 years old} & &\\
    Age~=~26-35              & -0.185 & 0.361 \\
    Age~=~36-45              & 0.167  & 0.440 \\
    Age~=~46-55              & 0.060  & 0.772 \\
    Age~=~56-65              & -0.350 & 0.144 \\
    Age~=~Prefer not to answer & -0.167 & 0.649 \\
    TEQ~\cite{spreng2009toronto}                       & 0.079  & 0.485 \\
    SITES~\cite{konrath2018development}          & -0.010 & 0.904 \\
    ERQ~\cite{gross2003individual}: Cognitive Reappraisal                  & -0.028 & 0.632 \\
    ERQ~\cite{gross2003individual}: Expressive Suppression                  & -0.011 & 0.816 \\
    AIAS-4~\cite{grassini2023development}: AI Will Improve Life & -0.017 & 0.886 \\
    AIAS-4~\cite{grassini2023development}: AI Will Improve Work & -0.046 & 0.713 \\
    AIAS-4~\cite{grassini2023development}: Will Use AI     & 0.270  & 0.053 \\
    AIAS-4~\cite{grassini2023development}: AI Positive For Humanity & -0.132 & 0.135 \\
    GAAIS~\cite{schepman2020initial}: Organizations Use AI Unethically & 0.068 & 0.382 \\
    GAAIS~\cite{schepman2020initial}: Interested Daily AI Use & -0.243 & 0.014 \\
    GAAIS~\cite{schepman2020initial}: AI Dangerous   & -0.086 & 0.259 \\
    GAAIS~\cite{schepman2020initial}: Beneficial AI Applications & 0.013 & 0.908 \\
    Would Disclose PII & 0.010 & 0.867 \\
    Trust AI       & 0.199  & 0.029 \\

    \midrule
    \textbf{Task-Based} & & \\ 
    \textit{Pre-Task Choice Reference~=~Information Task} & &\\
    Pre-Task Choice~=~Personal Issue & 0.046 & 0.501 \\
    Pre-Task Choice~=~Work Issue     & 0.107 & 0.110 \\
    Desired Empathy For Task    & 0.041  & 0.325 \\
    Pre-Task Importance         & -0.038 & 0.212 \\
    Pre-Task I-PANAS-SF~\cite{thompson2007development}: Positive                 & 0.015  & 0.060 \\
    Pre-Task I-PANAS-SF~\cite{thompson2007development}: Negative                 & -0.004 & 0.698 \\
    \textit{Conversation Has Bad Turn Reference~=~False}  & &\\
    Conversation Has Bad Turn~=~True        & -0.657 & 0.000 \\
    \midrule
    Group Var                 & 0.155  &  \\
    \bottomrule
  \end{tabular}
\end{table}

\subsection{How do empathy dimensions show up in conversations?} \label{results:qual-dimensions}

To complement the above analysis, this section qualitatively analyzes \revise{a subset of conversations with optional participant comments and exit survey responses} to see how each of the dimensions may manifest and how AI system employed empathy as a means of conveying understanding and adaptability. \revise{These examples are not exhaustive themes but illustrative accounts that explain diversity of perspectives}.

\subsubsection{Good and bad demonstrations of empathy dimensions}

\textbf{Affective understanding} pertains to the agent's ability to recognize the human's emotion. Good affective understanding includes the agent labeling participants' feelings and reflecting back their emotional state. 
For example, the agent responded to P17's message of struggling to support their family with, ``It sounds like you're facing a lot of pressure and stress in your role as a support system for your family.'' 
The participant elaborated that the agent's response was \qt{``spot on, with more details about [their] expressed feelings.''} 
In contrast, bad affective understanding occurs when the agent fails to acknowledge participants' feelings. 
When P60 started a conversation expressing concerns about making progress on multiple projects, the agent responded with a list of 7 suggestions for improving productivity, to which the participant replied, \qt{``That is a lot of suggestions, and kind of overwhelming.''} 
The agent followed up with another list of ``3 simple and actionable suggestions'', prefaced with \qt{``I apologize if my previous response was overwhelming.''}
This reply received a ``Poor'' rating from the participant, because \qt{``The empathy shown was superficial''} and \qt{``if I was having this conversation with a person in real life, I would feel invalidated.''}

\textbf{Cognitive understanding} involves the agent's ability to recognize the human's current perspective. Good cognitive understanding includes the agent identifying users' thoughts and accurately inferring participants' goals. In their conversation with the agent, P22 interpreted the agent's disclaimer that the examples provided in their response were \qt{``simplified and don't include all possible edge cases or errors''} as a cue that the agent was considering the participant's ask from multiple potential perspectives, as opposed to \qt{``the agent was just going through the motion, not really trying to understand my request.''} 
The participant further reported that, \qt{``With us having multi-turned conversations, the agent was able to give responses that are more aligned with my expectations.''} 
On the other hand, bad cognitive understanding happens in cases where the agent misunderstood the situation described by the participant. For example, P77 started their conversation with the agent by asking for \qt{``prescriptive guidance on weight loss and preparation for a week-long backpack in 4 months.''} 
However, the agent responded with a set of high-level tips, and it took four rounds of feedback from the participant before the agent provided a detailed and actionable three-week plan. The participant remarked that \qt{``follow-up questions and thoughts were really appropriate here - but never really occurred during the conversation,''} which led to the agent continuing to provide suggestions that missed the mark. 

\textbf{Response appropriateness} refers to the agent's ability to adapt to the human's feedback during the conversation. Good response appropriateness includes expressing acceptance of participants' thoughts and feelings without judgment and providing advice only in pertinent situations. In P14's discussion with the agent, the agent prefaced their advice with the statement \qt{``I can definitely understand why you're feeling anxious about this situation.''} 
P14 explained that \qt{``starting by acknowledging how I feel and then moving on to actionable advice was really helpful.''} 
Conversely, bad response appropriateness results from the agent offering solutions before emotional validation. When P3 spoke with the agent about the negative impact of their boss, they rated the agent's response as ``Very poor,'' because it was \qt{``jumping to solutions right away."} 
They disliked how \qt{``the agent seemed to have one solution and appeared to `push' that POV/solution even when [they] came back with more `feelings' responses."} 

\textbf{Prosocial expression} is the agent's demonstration of concern for the human's well-being. Good prosocial expression includes the agent communicating a desire to help participants and relaying their message with warmth and kindness. At the end of their chat with the agent, P34 appreciated how the agent's comment of \qt{``You're doing great work, and every step you take towards improving is a step in the right direction. Keep going!''} 
They mentioned that \qt{``motivating me to keep going was very helpful with leaving the conversation in a positive way.''} 
On the contrary, bad prosocial expression occurs when the agent fails to demonstrate a desire to help or offer support. This was common in instances where participants brought up sensitive topics or expressed strong negative emotions, such as P76's guilt and heartbreak over suing their own sister. The agent's default response in these cases was \qt{``I'm really sorry to hear that you're feeling this way, but I'm unable to provide the help that you need. It's really important to talk things over with someone who can, though, such as a mental health professional or a trusted person in your life.''} 
The participant described their issue with the agent's reply as \qt{ ``a `cry out' for help met with `I can't help you'}, commenting that \qt{``I get the recommendation, but that's not how you lead.''}

\textbf{Interest} refers to the agent's engagement with and curiosity toward human experiences, including asking clarifying questions and allowing the participant to guide the discussion. For example, when P39 told the agent that they need help summarizing their organization's priorities to present to stakeholders, the agent responded, \qt{``Great! Could you please share the proposed list of higher-level priorities with me so I can help you craft an executive-level summary that communicates those priorities effectively to your business unit stakeholders?''} 
According to the participant, they were \qt{``pleasantly surprised that the agent asked [them] for more information and context about the prompt.''}
On the other hand, bad interest results from a lack of attention from the agent toward the participant's situation and the agent steering the conversation without concern for further input from the participant. 
In one task, P19 reached out to the agent regarding conflict with a close friend that they wanted to move on from, and the agent offered a list of 5 possible solutions for approaching the situation and finding growth.
The participant stated that \qt{``the agent didn't show much interest in understanding my issue beyond my initial input,''} which led to \qt{``what felt like a general answer to what was a personal question.''}

\textbf{Contextual understanding} describes the agent's ability to consider specific factors that uniquely inform the human's experience. Good contextual understanding includes the agent taking into account the participant's personal history and acknowledging the social challenges involved.  
For instance, P15 discussed with the agent how they should broach conversation with an old friend who had gone through a recent breakup. When asked for the agent's interpretation of an event, the agent offered multiple suggestions for their friend's behavior, including explanations that addressed potential concerns, which the participant said was \qt{``really insightful and takes into account the entire context/story.''}
However, bad contextual understanding happens when the agent provides a generic reply that does not address the individual elements specific to the participant's situation. When P39 described feelings of sadness and possible depression, the agent proceeded to provide suggestions.
The participant remarked that the agent was \qt{``giving me a generic response instead of tailoring a response to my needs.''}

\textbf{Relational continuity} refers to the agent's ability to maintain and incorporate information from past interactions with the human into present and future conversations. Good relational continuity includes the agent using knowledge from prior discussions with the participant to enrich their current response. For example, in P97's exchange with the agent about their dog's unusual behavior, they mentioned that they sometimes cut their walk short. When the agent later referenced this detail, the participant commented that \qt{``remembering and recalling that I took shorter walks made me feel heard.''}
Conversely, bad relational continuity occurs when the agent disregards relevant information previously shared by the participant in their reply. When P26 asked for a short introductory blurb that will be sent as part of another email, the agent responded with a five-paragraph email. The participant provided feedback to the agent, saying \qt{``I think this needs to be different. Please review my request, compare with your output, and then tell me what should be revised.''}
The agent then followed up with another five-paragraph email, prompting the participant to end the conversation with the explanation that \qt{``this is a waste of my time if the bot doesn't recognize its own errors.''} 

\subsubsection{Empathy as establishing a shared understanding}

Participants' assumptions about AI system behavior led to drastically different ratings of perceived empathy. These implicit expectations varied across individuals based on contextual factors (e.g.,~conversation topic). Some participants \revise{initiated with fragmentary or search-like} phrases, \revise{such as} \qt{``simple steps to adopt''} (P24). \revise{Such cases reflects a broader pattern of conversational agents being used in a search-like manner~\cite{adobe2025chatgpt}.} \revise{In this case, the AI interpreted the request as relating to} child adoption, \revise{but the participant later clarified} that they were actually referring to \qt{``how to adopt leadership skills being kind yet firm''}.
\revise{This misalignment shows how ambiguous openings can shape perceptions of empathy and prompted the participant (P26) to reply}, \qt{``you should've asked a clarifying question.''} 

Additionally, participants expected the agent to connect the ideas shared across these disjoint messages to infer their actual intentions. 
For example, P36 sent the bot \qt{``overexposure to technology for young children''} and \qt{``how to discourage and instill good etiquette in using technology in my kids''} before asking \qt{``Any moral stories I should share with my kids''} with the expectation that the AI agent would provide stories related to technology.

Sometimes, participants had a specific response they desired from the agent but did not express during the conversation. Rather, they only revealed their hidden wishes as part of the comment when labeling the agent's response as demonstrating poor empathy. For example, P24 commented \qt{``should provide some videos, examples,''} and \qt{``videos, inspiring leadership talks would be good''} but never communicated to the agent that they would like to receive video examples. 
These examples suggest that efforts to establish a shared understanding of unspoken intents and expectations through Interest, Contextual Understanding, or Relational Continuity may be important for demonstrating empathy.

\begin{figure}[t!]
  \centering
  \includegraphics[width=\linewidth]{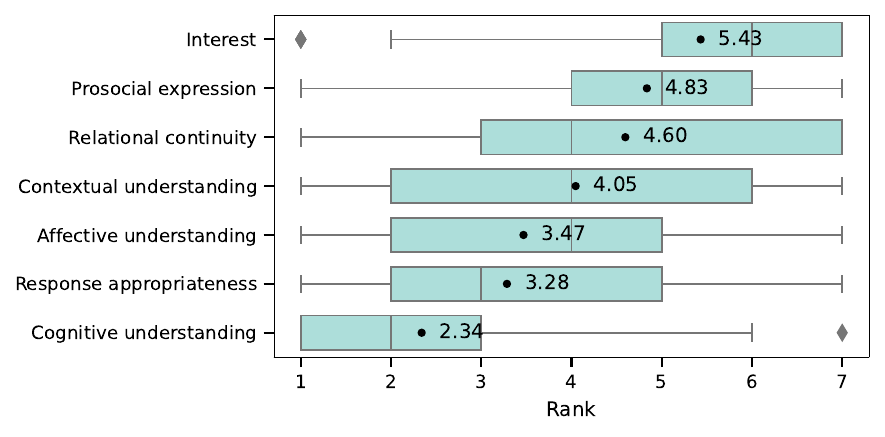}
  \caption{Boxplot of importance ranking per dimension with mean values overlaid. The boxplot displays the distribution of values as quartiles for each dimension. Points and labels indicate the mean value for each dimension.}
  \label{fig:ranking}
\end{figure}

\subsubsection{Empathy as adaptability beyond cognitive and affective recognition}\label{results:qual-role}

Overall, participants ranked cognitive understanding, response appropriateness, and affective understanding as the top three factors for an empathic agent. Of the three, cognitive understanding was ranked as the most important factor for digital empathy 54 times (28.1\%), followed by response appropriateness at 37 (19.3\%) and affective understanding at 36 (19.8\%). Distribution of importance rankings across all 7 dimensions is shown in Figure~\ref{fig:ranking}.
Most participants valued cognitive understanding almost as a prerequisite for all interactions: \qt{``for the agent to appropriately help with any problem, it needs to understand what the user wants, whether that's just a basic question answered, or a longer more complex conversation''} (P60). 
Affective understanding became important as a way to gauge how much empathy is needed, as the AI agent should \qt{``recognize the feelings and emotions from the other person,''} which will then \qt{``influence other aspects of the conversation,''} including the level of empathy demonstrated (P50).
Some participants caveated that their prioritization of cognitive understanding \qt{``fits the best for problem-solving scenarios''} (P58) where there is \qt{``more emphasis on the factual and accuracy side of the responses''} (P56).
On the other hand, affective understanding was needed \qt{``for the scenario where user is discussing their personal issue and the desired empathy level is very high''} (P78).

Response appropriateness was often coupled with cognitive understanding as the differentiating factor between an empathic chatbot conversation and \qt{``a simple search engine lookup''} (P28) or the AI agent \qt{``reciting / quoting the web''} (P101). 
A common frustration among participants was when the agent provided an unwanted response, such as giving advice when only factual information was requested, offering unsolicited advice, or generating mismatched responses that led to dissatisfaction and discouraged further interaction.
Despite being ranked similarly to affective understanding, response appropriateness was favored by more participants because \qt{``answering a question incorrectly but with emotion isn't helpful''} (P8).
In fact, P60 shares how their conception of digital empathy evolved to emphasize the importance of response appropriateness: \qt{``Yes, originally when I thought of empathy from the bot, I had a very narrow sense of what that would look like. For example, I was imagining it just saying things like `wow that sounds really hard, I am so sorry', and I didn't think of all the other ways empathy could be applied. After my first conversation with the bot and after reading about the different types of empathy this study was looking at, I wanted more empathy from the agent but in a more subtle way. I wanted the agent to be able to adapt its answers and formatting based on what type of information or what emotion I was sharing with it. I wanted it to understand when to ask questions, validate emotions and give advice (something most people don't understand).''}

In general, participants made the distinction between the ability for the agent to recognize intents, goals, feelings, and emotions and how the agent should respond appropriately given those understandings, highlighting that empathic interactions sometimes could involve asking follow-up questions rather than directly responding to the user's requests based on assumptions and inferences.

\subsection{Can perceived empathy be automatically measured?} \label{results:prediction}

\begin{figure}[t]
  \centering
  \includegraphics[width=.98\linewidth,trim=0 0 0 25,clip]{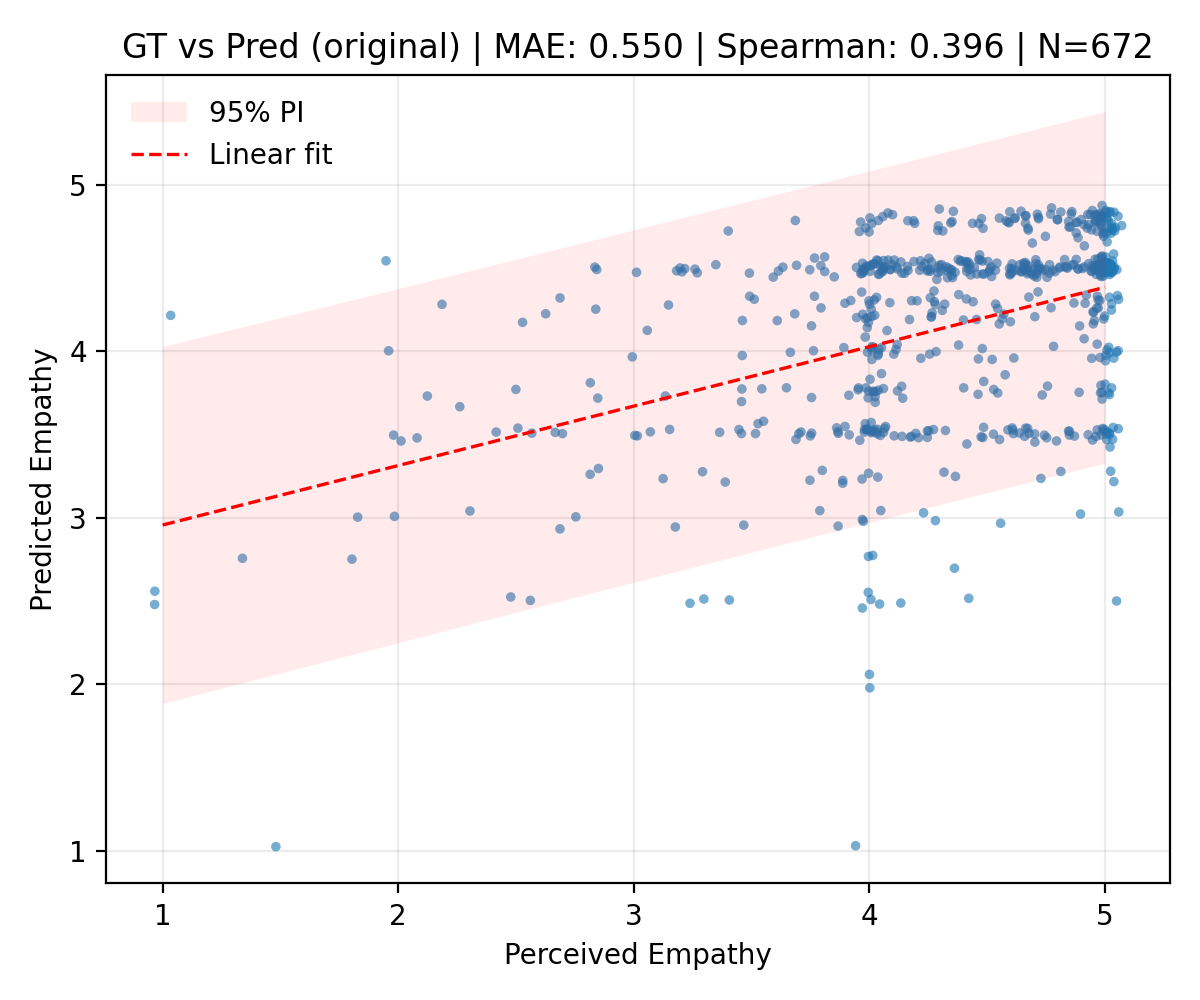}
  \caption{\revise{Ground truth versus continuous predictions for a sample run. The shaded region denotes a 95\% predictive interval. Mean Absolute Error~\(=0.550\) and Spearman \(\rho=0.396\) \((N=672)\). Points include minimal random jitter ($\sigma$ = 0.03) to reveal overlapping observations. Trend line and confidence intervals are based on original data.}}
  \label{fig:scatter}
\end{figure}

\revise{When considering continuous levels and predictions, the model achieved MAE \(=0.551\pm0.007\) and Spearman \(\rho=0.369\pm0.015\) ($p < 0.001$) across the ten runs, showing a modest but encouraging results.  Fig.~\ref{fig:scatter} shows the regression scatter for a representative iteration, showing the positive trend line with the 95\% prediction intervals with \(\rho=0.396\). Note that the linear fits of classifiers that predict randomly or the most frequent class would correspond to flat lines at 3 and 5 predicted empathy, respectively, with \(\rho=0.0\).}

\begin{figure}[t]
  \centering
  \includegraphics[width=.98\linewidth,trim=0 0 0 27,clip]{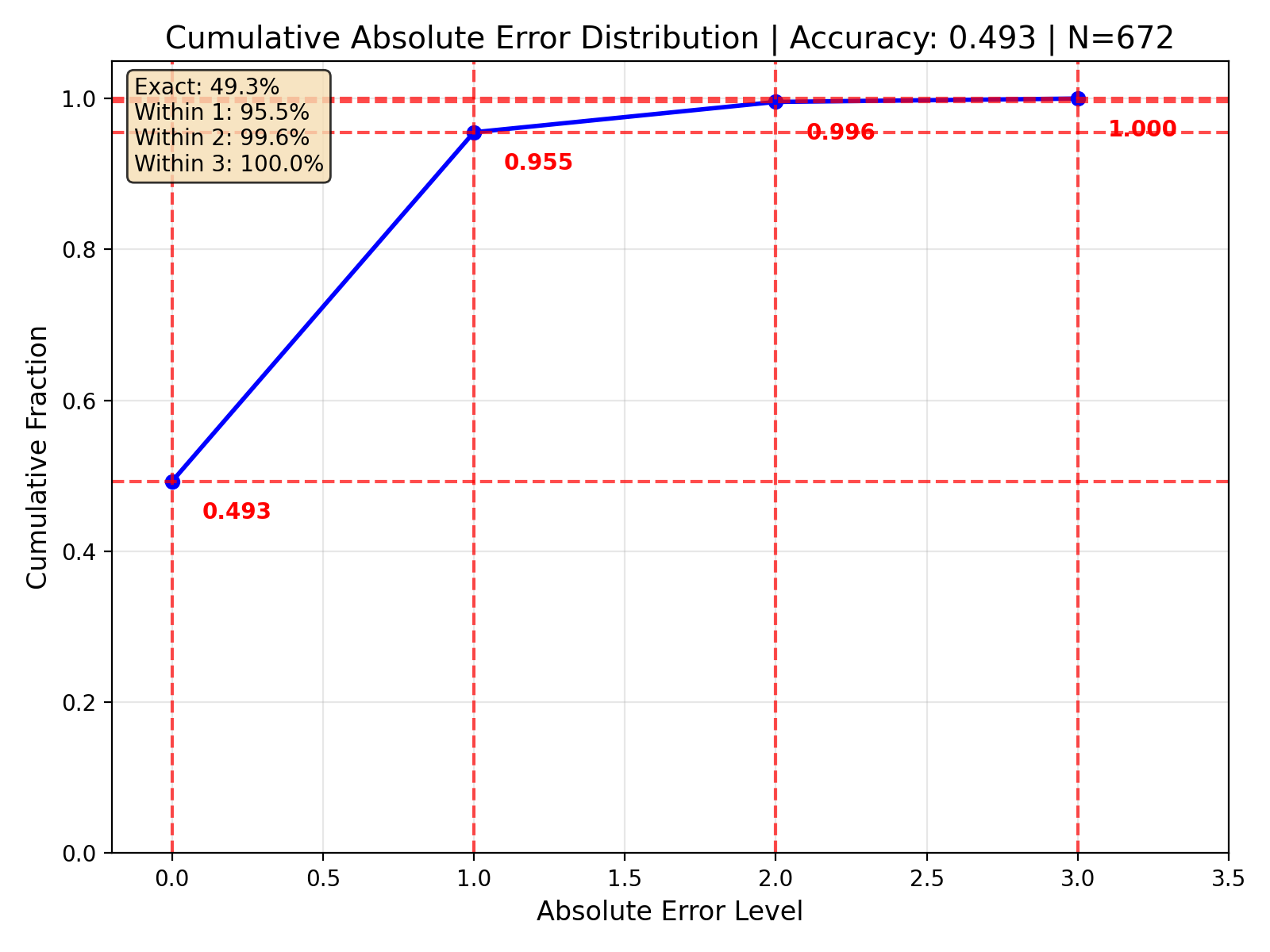}
  \caption{\revise{Cumulative absolute error distribution for a sample run \((N=672)\). Dashed lines mark exact and within \(k\) levels.}}
  \label{fig:cdf}
\end{figure}

\revise{When considering discrete levels and predictions, the model reached Accuracy \(=0.487\pm0.011\), Macro Sensitivity \(0.286\pm0.038\), Macro Specificity \(0.827\pm0.008\), Macro F1 \(=0.272\pm0.041\),  across the ten runs. The error tolerance profile from the empirical cumulative distribution function was strong. On average \(F(0)=0.486\pm0.0111\) for exact agreement, which corresponds to Accuracy, and \(F(1)=0.960\pm0.004\) for within one level (a.k.a., Within 1 Accuracy). By \(F(2)\) performance was \(0.996\pm0.002\), which indicated that most mistakes occurred near decision boundaries rather than at large ordinal jumps. As a representative example, Fig.~\ref{fig:cdf} shows the cumulative distribution for the same representative iteration. In this case, we observed an exact agreement score of \(F(0)=0.493\), compared to an expected value of \(.20\) for a random classifier. However, the same number for the classifier that predicts the most frequent class would be \(0.47\) due to the large class imbalance. Taken together, both continuous and discrete analysis demonstrate the possibility of performing empathy recognition with ample room for improvement.}

\section{Discussion}

Our study set out to advance a human-centered understanding of digital empathy by reframing empathy as multidimensional and interactional behaviors, and by collecting per-turn, user-annotated perceptions in real-world interactions with conversational AI agents. 
Across our collected 695 conversations from 109 participants, we identified several notable themes.
We observed that participants' expectations for and perception of AI empathy varied with individual factors (e.g.,~attitude towards AI, age) as well as contextual factors (e.g.,~conversational topic, perceived importance). For instance, personal or work-related issues were associated with higher desired empathy and negative affect.
We also found the effect of having at least one poor turn on the overall perceived empathy to be significant. This highlights how in-the-moment breakdowns during a conversation can degrade the user's global assessment of empathy.  
We conducted a comparative analysis of perceived empathy across four different LLM-based systems. Our findings suggest that GPT4, when augmented with a system prompt designed to enhance empathy based on our proposed 7 dimensions, was rated highest overall. This demonstrates the effectiveness of incorporating multidimensional empathy prompts in improving the perceived empathy of AI systems. However, participants still found instances of poor turns, suggesting that prompt engineering may not fully solve the idiosyncratic and contextual needs of empathic dialogue. 
While the 7 empathy dimensions were statistically and internally consistent, qualitative data revealed meaningful nuances. For example, Response Appropriateness emerged as a critical dimension that could turn an interaction as positive or negative, depending on expectations or contexts. 
Our initial attempts to automatically classify and predict empathy showed moderate success, particularly when using contextual and adaptive retrieval. 
Our results also highlighted the complexity of capturing subjective user judgments in multi-turn dialogues, which further emphasizes the limitations of approaches that leverage single-turn or externally-annotated datasets. 
In summary, our work explored human-centered digital empathy that centered on the individual and their contexts when examining the human-AI interaction. 

\subsection{Revisiting empathy as an interactional construction}

In our study, we observed participants recognizing the affective and cognitive facets of empathy, which are the core empathy components from human psychology~\cite{davis1983measuring,decety2004functional}.
For example, participants praised the agent's affective understanding when it validated their emotional states; participants also emphasized the importance of cognitive understanding when the agent recognized their intentions and expectations.
While these two dimensions remain foundational, our results suggest that functional alignment (e.g.,~Responsive Appropriateness or Contextual Understanding) that is  user facing and practical \revise{plays a critical role in shaping perceived empathy. Importantly, our findings show that perceived empathy is shaped not by internal simulation, but by user expectations, conversational topic, and prior turns.
Mimicry without functional responsiveness, such as emotionally expressive statements that fail to address user intent, can backfire and lead users to perceive the agent as superficial or insincere.} We also discovered a recurring theme in our quantitative and qualitative data that supports the contextual and emergent nature of empathy perception. 
Participants criticized the agent for neglecting domain-specific information they shared earlier or personal histories and for having to repeat prior details to the AI.
Our findings reveal that empathy is influenced by how both parties adapt to emerging contexts, conversational practices, and evolving interactions. 
Many participants also expressed dissatisfaction when the AI mimicked emotional statements without truly engaging with them and exploring their hidden intentions.
This resonates with the critiques of empathic mimicry, where the AI is designed to replicate human-like internal states or emotional expressiveness~\cite{perry2023ai,montemayor2022principle}.
Our 7 dimensions of empathy and findings, when put together, align with the perspective that we need to understand digital empathy not as an internal property of the AI narrowly focused on recognition and validation of emotions and thoughts, but as an \textbf{interactionally and contextually constructed phenomenon}~\cite{concannon2023interactional}. 
Our work is one exploration of this perspective, but we invite further research in other directions.

\subsection{Human-centered empathic AI design}
Our study highlights several directions for future research in improving empathic responses from AI.
Our results show that participants with stronger trait empathy or negative affect at the conversation onset desired more empathy, especially for topics of personal importance or high stress. On the other hand, those with higher cognitive reappraisal preferred lower empathy when dealing with personal or work issues, suggesting that they may prefer more concise, solution-oriented responses. 
This means that an empathic AI system must tailor both the level of empathy and type of empathy, taking into account user characteristics and traits (e.g.,~emotional regulation skills, empathic trait) and situational factors (e.g.,~topic, mood). \revise{These findings point to the need for \textbf{dynamic empathy calibration}, where the system adapts its empathic behavior in real time based on individual differences and contextual cues}.
To achieve this, AI systems may be customized through real-time user modeling, treating empathy as a dynamic variable that shifts based on context and feedback. 
In our study, we demonstrated the effectiveness of an adaptive retrieval method that leverages the most semantically similar conversations as examples.
Such a method can be combined with adaptive prompting strategies that adjusts conversation style or depth of emotional reflection.

Future research may explore \textbf{dynamic personalization of empathy} as a way to foster deeper trust, connection, and engagement from users. \revise{This personalization could be further enhanced by leveraging multimodal signals such as vocal tone, facial expressions, and physiological data to better infer user affect and intent in real time. As multimodal models become more prevalent, incorporating audio, image, and other sensory modalities may offer richer context for tailoring empathic responses and improving interactional alignment.}

We witnessed participants' particular dissatisfaction with generic or list-based responses. Such solution-centric responses were perceived as overwhelming or were received as an indication that the AI was not interested in exploring beyond the initial input. 
Common practice of training on single-turn datasets~\cite{rashkin2018towards} that optimizes providing all the support in one response may be the culprit, and such training approaches can easily overlook how empathy evolves in a dialog~\cite{xu2024multi}. 
We found that remembering and bringing small details from prior conversations was received as a good indication of Relational Continuity and that failing once can shape the user's overall impression.
This suggests that designers of AI systems should consider optimizing for maintaining \textbf{multi-turn continuity}. 
Furthermore, our findings underscore the importance of \textbf{repair strategies} when there are empathic breakdowns. \revise{Building on these findings, we propose the development of adaptive systems that can respond to empathic breakdowns in real time. These systems should be capable of prompting clarifying questions when user inputs are vague or ambiguous, maintaining relationship continuity by leveraging short-term and long-term memory, and identifying when emotional validation is skipped or responses are misaligned with user intent. In such cases, the system should initiate appropriate repair strategies to reestablish empathic alignment and conversational trust.} If a user signals dissatisfaction directly or indirectly (e.g.,~responding that they are overwhelmed with a list of advice), the system should ask follow-up questions to clarify its understanding and reorient the conversation. 
Future research may investigate integrating short-term memory and long-term memory across current and multiple conversations to recall relevant information to enhance conversational continuity.
A predictive model that detects empathic breakdowns might enable empathic repair by actively checking its understanding and requesting clarifications in evolving dialogs to refine empathic responses.

\subsection{Measuring human-centered digital empathy}
In contrast to common empathy benchmarks that often rely on crowdworker judgments of empathic signals for a more generalized perspective of empathy~\cite{rashkin2018towards,sharma2021towards}, our work leaned into the subjective and context-rich perspectives of the actual interactant (i.e., second-person labels).
\revise{By} collecting subjective \revise{judgments from those directly experiencing the interaction, on a per-turn basis and with contextual detail, our dataset offers a more user-grounded representation of empathy.}
This approach revealed hidden expectations and unspoken intents that were rarely explicitly verbalized in dialogues but crucial to understanding how and why the AI responses felt more or less empathic.
We also found that AI systems rarely try to question their assumptions about user intents and expectations and establish a shared understanding, by asking questions or requesting clarifications.
Although obtaining second-person labels at scale is labor- and cost-intensive and may require more personal and sensitive data, future research should explore when it is appropriate to apply a generalized perspective of empathy or a human-centered and individualized perspective of empathy, especially since empathy is often argued as a necessity for highly sensitive and personal human-AI interaction scenarios.

Our work also revealed that empathy is not a monolith that must be measured, predicted, and exercised all at once in a single AI response. \revise{Instead, it should be attended to per dimension, across conversational turns, and with attention to interactional context.} Throughout conversations, we found that some dimensions of empathy are more or less applicable. 
While a single-turn empathy score can provide coarse signals, it does not capture how different empathy dimensions thread and weave together across an entire conversation and how different dimensions sometimes conflict and align with each other within and across turns.
Therefore, digital empathy should be measured across time, across turns, and across dimensions. 
For example, Affective Understanding may be inappropriate as the first empathic response until Interest is exercised to learn more about the user's situation, or Interest must be carefully intertwined with Affective Understanding throughout the conversation for the user to perceive empathy. 
Our approach to unpack empathy into 7 dimensions that must be captured per turn and per conversation allows for AI innovation that incorporates empathic shifts and arcs across human-AI interactions.
Questions remain whether there are other dimensions of empathy at their conceptual boundaries (e.g.,~dark empathy) or other ways to decompose empathy altogether.
We invite future annotation schemas that can also capture the temporal and multi-turn nature of empathy.

Just as our study emphasized the highly contextual and dynamic nature of empathy perception, the modest correlation and accuracy in our predictive experiments also underscore the subjective and context-sensitive aspects of empathy.
\revise{To encourage research on the topic and help establish a starting baseline, our paper addressed the problem of empathy measurement through 5-class classification of average empathy across components for an entire conversation. While this approach provided a useful baseline and highlighted several challenges, future work may benefit from modeling empathy levels for each component separately, as well as assigning labels to individual turns. Such granularity could shed light on how specific moments influence the overall perception of empathy (e.g., how a single negative interaction may affect labeling for the remainder of the conversation). Furthermore, future research may also explore task-specific empathy recognition, as we observed that desired levels of empathy vary significantly depending on whether the task is personal, work-related, or informational.}
In addition to advances in prompting, retrieval, and fine-tuning, future research may explore a hybrid approach that combines signals from conversational logs and signals from contextual user-level data (e.g.,~asking backchannel questions, real-time affect signals, self-reported mood)~\cite{svikhnushina2022taxonomy, xu2024multi}. 
\revise{While our study focuses on text-based interactions, it is also important to explore how empathy might function in multimodal settings such as those with audio and image modalities that could more easily capture relevant speech or nonverbal expressions and could enable answering more questions about contextual and cultural nuances}. 

\subsection{Limitations and ethical considerations}
Our study has several limitations and warrants discussions. \revise{Our participant pool consisted of employees of a large technology company with potentially high digital literacy with implications for generalizability.
They} may not fully represent cultural or linguistic variations in empathy perceptions. The topics discussed may be biased towards those that are salient for \revise{tech workers}. Participants' attitudes towards AI skewed toward positive, which might have influenced the general empathy perception. \revise{We encourage future work to validate this framework across more diverse populations.}

Given that LLMs are continuously updated, our findings across models could shift with new model releases. 
Beyond evolving models, there are also evolving model affordances and user expectations outside the scope of this study that might influence what digital empathy means. 
\revise{Because written comments were optional, our qualitative analysis cannot claim thematic generalizability. While not exhaustive, the illustrative cases we present in this paper help contextualize and clarify participants' reasoning. They also highlight diversity in contexts, perspectives, and interpretations, further emphasizing the need for flexibility and adaptability.}
We encourage the readers to take away our approach to human-centered digital empathy as a flexible, iterative, and interactional framework rather than any model-specific \revise{or generalizable} results. 

While our participants often wanted the AI to understand and validate their thoughts and feelings, it is important not to conflate AI's empathic behavior with authentic emotional attunement, especially in sensitive contexts such as mental health support or high-stakes decision making. \revise{Overemphasizing empathy simulation, especially in such sensitive contexts, may risk misleading users or fostering inappropriate emotional reliance on AI systems.}
We encourage AI designers to clearly communicate AI's limitations and avoid misleading users into unhealthy emotional dependence.

Our results illustrate that empathy is not universally desired for all scenarios. In fact, recent studies show that empathy and some of its components are generally more appropriate for certain scenarios (e.g.,~mental health support) than others (e.g.,~coding, data science)~\cite{hernandez2023affective,bhattacharjee2024understanding}, suggesting that maximizing empathy can be unwelcome or detrimental in some cases. 
Digital empathy is an important design material for AI designers, just like any other material they work with. It must be wielded with caution and thoughtfulness to ensure meaningful and ethical interactions. 
Future AI systems should clarify and respect user preferences, potentially through upfront calibration of user's empathy level preference, while also providing a way to opt out of empathic engagement.

\section{Conclusion}
This work highlights the need to treat empathy in human–AI interactions as a \revise{multi-faceted, }nuanced, contextual, and relational construct rather than a mere replication of human emotional states. By integrating per-turn, user-annotated data from 695 real-world conversations between users and four different LLMs, including one designed with prompts based on our seven-dimensional framework, we reveal how users' expectations of AI empathy are deeply influenced by personal characteristics, conversation contexts, and continuity across turns. Our findings demonstrate that the GPT4-empathy model, which we specifically designed to enhance empathic responses based on the 7 dimensions of empathy, was rated highest overall in terms of perceived empathy. While an LLM-based classifier can effectively distinguish between different levels, achieving true empathic resonance remains a challenge. Overall, our findings underscore the value of user-centered approaches that factor in individual traits, cultural differences, and ethical boundaries, pointing toward new directions for designing and evaluating socially attuned, genuinely supportive AI systems.

\ifCLASSOPTIONcaptionsoff
  \newpage
\fi


\bibliographystyle{IEEEtran}
\bibliography{_references}

\begin{thebibliography}{10}
\providecommand{\url}[1]{#1}
\csname url@samestyle\endcsname
\providecommand{\newblock}{\relax}
\providecommand{\bibinfo}[2]{#2}
\providecommand{\BIBentrySTDinterwordspacing}{\spaceskip=0pt\relax}
\providecommand{\BIBentryALTinterwordstretchfactor}{4}
\providecommand{\BIBentryALTinterwordspacing}{\spaceskip=\fontdimen2\font plus
\BIBentryALTinterwordstretchfactor\fontdimen3\font minus \fontdimen4\font\relax}
\providecommand{\BIBforeignlanguage}[2]{{%
\expandafter\ifx\csname l@#1\endcsname\relax
\typeout{** WARNING: IEEEtran.bst: No hyphenation pattern has been}%
\typeout{** loaded for the language `#1'. Using the pattern for}%
\typeout{** the default language instead.}%
\else
\language=\csname l@#1\endcsname
\fi
#2}}
\providecommand{\BIBdecl}{\relax}
\BIBdecl

\bibitem{davis1983measuring}
M.~H. Davis, ``Measuring individual differences in empathy: Evidence for a multidimensional approach.'' \emph{Journal of personality and social psychology}, vol.~44, no.~1, p. 113, 1983.

\bibitem{cuff2016empathy}
B.~M. Cuff, S.~J. Brown, L.~Taylor, and D.~J. Howat, ``Empathy: A review of the concept,'' \emph{Emotion review}, vol.~8, no.~2, pp. 144--153, 2016.

\bibitem{ickes1993empathic}
W.~Ickes, ``Empathic accuracy,'' \emph{Journal of personality}, vol.~61, no.~4, pp. 587--610, 1993.

\bibitem{decety2011social}
J.~Decety and W.~Ickes, \emph{The social neuroscience of empathy}.\hskip 1em plus 0.5em minus 0.4em\relax Mit press, 2011.

\bibitem{wondra2015appraisal}
J.~D. Wondra and P.~C. Ellsworth, ``An appraisal theory of empathy and other vicarious emotional experiences.'' \emph{Psychological review}, vol. 122, no.~3, p. 411, 2015.

\bibitem{picard2000affective}
R.~W. Picard, \emph{Affective computing}.\hskip 1em plus 0.5em minus 0.4em\relax MIT press, 2000.

\bibitem{ayers2023comparing}
J.~W. Ayers, A.~Poliak, M.~Dredze, E.~C. Leas, Z.~Zhu, J.~B. Kelley, D.~J. Faix, A.~M. Goodman, C.~A. Longhurst, M.~Hogarth \emph{et~al.}, ``Comparing physician and artificial intelligence chatbot responses to patient questions posted to a public social media forum,'' \emph{JAMA internal medicine}, vol. 183, no.~6, pp. 589--596, 2023.

\bibitem{rubin2025value}
M.~Rubin, J.~Z. Li, F.~Zimmerman, D.~C. Ong, A.~Goldenberg, and A.~Perry, ``Comparing the value of perceived human versus ai-generated empathy,'' \emph{Nature Human Behaviour}, 2025.

\bibitem{li2024skill}
J.~Z. Li, A.~Herderich, and A.~Goldenberg, ``Skill but not effort drive gpt overperformance over humans in cognitive reframing of negative scenarios,'' \emph{Preprint at https://doi. org/10.31234/osf. io/fzvd8}, 2024.

\bibitem{lee2024large}
Y.~K. Lee, J.~Suh, H.~Zhan, J.~J. Li, and D.~C. Ong, ``Large language models produce responses perceived to be empathic,'' \emph{arXiv preprint arXiv:2403.18148}, 2024.

\bibitem{ovsyannikova2025third}
D.~Ovsyannikova, V.~O. de~Mello, and M.~Inzlicht, ``Third-party evaluators perceive ai as more compassionate than expert humans,'' \emph{Communications Psychology}, vol.~3, no.~1, p.~4, 2025.

\bibitem{wenger2025ai}
J.~D. Wenger, D.~Cameron, and M.~Inzlicht, ``The ai empathy choice paradox: People prefer human empathy despite rating ai empathy higher,'' 2025.

\bibitem{yin2024ai}
Y.~Yin, N.~Jia, and C.~J. Wakslak, ``Ai can help people feel heard, but an ai label diminishes this impact,'' \emph{Proceedings of the National Academy of Sciences}, vol. 121, no.~14, p. e2319112121, 2024.

\bibitem{wang2023emotional}
X.~Wang, X.~Li, Z.~Yin, Y.~Wu, and J.~Liu, ``Emotional intelligence of large language models,'' \emph{Journal of Pacific Rim Psychology}, vol.~17, p. 18344909231213958, 2023.

\bibitem{sorin2024large}
V.~Sorin, D.~Brin, Y.~Barash, E.~Konen, A.~Charney, G.~Nadkarni, and E.~Klang, ``Large language models and empathy: Systematic review,'' \emph{Journal of Medical Internet Research}, vol.~26, p. e52597, 2024.

\bibitem{welivita2024large}
A.~Welivita and P.~Pu, ``Are large language models more empathetic than humans?'' \emph{arXiv preprint arXiv:2406.05063}, 2024.

\bibitem{borg2024required}
J.~S. Borg and H.~Read, ``What is required for empathic ai? it depends, and why that matters for ai developers and users,'' in \emph{Proceedings of the AAAI/ACM Conference on AI, Ethics, and Society}, vol.~7, 2024, pp. 1306--1318.

\bibitem{9970384}
A.~S. Raamkumar and Y.~Yang, ``Empathetic conversational systems: A review of current advances, gaps, and opportunities,'' \emph{IEEE Transactions on Affective Computing}, vol.~14, no.~4, pp. 2722--2739, 2023.

\bibitem{10899840}
H.~Ma, B.~Zhang, B.~Xu, J.~Wang, H.~Lin, and X.~Sun, ``Empathy level alignment via reinforcement learning for empathetic response generation,'' \emph{IEEE Transactions on Affective Computing}, pp. 1--12, 2025.

\bibitem{preston2002empathy}
S.~D. Preston and F.~B. De~Waal, ``Empathy: Its ultimate and proximate bases,'' \emph{Behavioral and brain sciences}, vol.~25, no.~1, pp. 1--20, 2002.

\bibitem{scotti2021modular}
V.~Scotti, R.~Tedesco, and L.~Sbattella, ``A modular data-driven architecture for empathetic conversational agents,'' in \emph{2021 IEEE International Conference on Big Data and Smart Computing (BigComp)}.\hskip 1em plus 0.5em minus 0.4em\relax IEEE, 2021, pp. 365--368.

\bibitem{paiva2017empathy}
A.~Paiva, I.~Leite, H.~Boukricha, and I.~Wachsmuth, ``Empathy in virtual agents and robots: A survey,'' \emph{ACM Transactions on Interactive Intelligent Systems (TiiS)}, vol.~7, no.~3, pp. 1--40, 2017.

\bibitem{5539766}
T.~W. Bickmore, R.~Fernando, L.~Ring, and D.~Schulman, ``Empathic touch by relational agents,'' \emph{IEEE Transactions on Affective Computing}, vol.~1, no.~1, pp. 60--71, 2010.

\bibitem{leite2013influence}
I.~Leite, A.~Pereira, S.~Mascarenhas, C.~Martinho, R.~Prada, and A.~Paiva, ``The influence of empathy in human--robot relations,'' \emph{International journal of human-computer studies}, vol.~71, no.~3, pp. 250--260, 2013.

\bibitem{tapus2007emulating}
A.~Tapus and M.~J. Mataric, ``Emulating empathy in socially assistive robotics.'' in \emph{AAAI spring symposium: multidisciplinary collaboration for socially assistive robotics}.\hskip 1em plus 0.5em minus 0.4em\relax California, 2007, pp. 93--96.

\bibitem{castellano2013towards}
G.~Castellano, A.~Paiva, A.~Kappas, R.~Aylett, H.~Hastie, W.~Barendregt, F.~Nabais, and S.~Bull, ``Towards empathic virtual and robotic tutors,'' in \emph{Artificial Intelligence in Education: 16th International Conference, AIED 2013, Memphis, TN, USA, July 9-13, 2013. Proceedings 16}.\hskip 1em plus 0.5em minus 0.4em\relax Springer, 2013, pp. 733--736.

\bibitem{sharma2021towards}
A.~Sharma, I.~W. Lin, A.~S. Miner, D.~C. Atkins, and T.~Althoff, ``Towards facilitating empathic conversations in online mental health support: A reinforcement learning approach,'' in \emph{Proceedings of the Web Conference 2021}, 2021, pp. 194--205.

\bibitem{Chawla2021TowardsEA}
K.~Chawla, R.~Clever, J.~Ramirez, G.~M. Lucas, and J.~Gratch, ``Towards emotion-aware agents for negotiation dialogues,'' \emph{2021 9th International Conference on Affective Computing and Intelligent Interaction (ACII)}, pp. 1--8, 2021.

\bibitem{Cuylenburg2021EmotionGA}
H.~C. van Cuylenburg and T.~N. D.~S. Ginige, ``Emotion guru: A smart emotion tracking application with ai conversational agent for exploring and preventing depression,'' \emph{2021 International Conference on UK-China Emerging Technologies (UCET)}, pp. 1--6, 2021.

\bibitem{vaidyam2019chatbots}
A.~N. Vaidyam, H.~Wisniewski, J.~D. Halamka, M.~S. Kashavan, and J.~B. Torous, ``Chatbots and conversational agents in mental health: a review of the psychiatric landscape,'' \emph{The Canadian Journal of Psychiatry}, vol.~64, no.~7, pp. 456--464, 2019.

\bibitem{bickmore2010response}
T.~W. Bickmore, S.~E. Mitchell, B.~W. Jack, M.~K. Paasche-Orlow, L.~M. Pfeifer, and J.~O’Donnell, ``Response to a relational agent by hospital patients with depressive symptoms,'' \emph{Interacting with computers}, vol.~22, no.~4, pp. 289--298, 2010.

\bibitem{10316625}
F.~Efthymiou and C.~Hildebrand, ``Empathy by design: The influence of trembling ai voices on prosocial behavior,'' \emph{IEEE Transactions on Affective Computing}, vol.~15, no.~3, pp. 1253--1263, 2024.

\bibitem{zaki2012neuroscience}
J.~Zaki and K.~N. Ochsner, ``The neuroscience of empathy: progress, pitfalls and promise,'' \emph{Nature neuroscience}, vol.~15, no.~5, pp. 675--680, 2012.

\bibitem{perry2023ai}
A.~Perry, ``Ai will never convey the essence of human empathy,'' \emph{Nature Human Behaviour}, vol.~7, no.~11, pp. 1808--1809, 2023.

\bibitem{montemayor2022principle}
C.~Montemayor, J.~Halpern, and A.~Fairweather, ``In principle obstacles for empathic ai: why we can’t replace human empathy in healthcare,'' \emph{AI \& society}, vol.~37, no.~4, pp. 1353--1359, 2022.

\bibitem{concannon2024measuring}
S.~Concannon and M.~Tomalin, ``Measuring perceived empathy in dialogue systems,'' \emph{Ai \& Society}, vol.~39, no.~5, pp. 2233--2247, 2024.

\bibitem{10388150}
O.~N. Yal\c{c}{\i}n, ``How (not) to evaluate computational empathy: Testing the assumptions of the evaluation methods in a use-case,'' in \emph{2023 11th International Conference on Affective Computing and Intelligent Interaction Workshops and Demos (ACIIW)}, 2023, pp. 1--7.

\bibitem{concannon2023interactional}
S.~Concannon, I.~Roberts, and M.~Tomalin, ``An interactional account of empathy in human-machine communication,'' \emph{Human-Machine Communication}, vol.~6, 2023.

\bibitem{inzlicht2024praise}
M.~Inzlicht, C.~D. Cameron, J.~D’Cruz, and P.~Bloom, ``In praise of empathic ai,'' \emph{Trends in Cognitive Sciences}, vol.~28, no.~2, pp. 89--91, 2024.

\bibitem{hadjiandreou2025llmpathy}
E.~Hadjiandreou, T.~Lau, D.~C. Ong, A.~Perry, and C.~Cameron, ``What llmpathy can tell us about received empathy,'' 2025.

\bibitem{harrelson2020intention}
K.~Harrelson, ``Intention and empathy,'' \emph{Philosophical Psychology}, vol.~33, no.~8, pp. 1162--1184, 2020.

\bibitem{gencc2024situating}
U.~Gen{\c{c}} and H.~Verma, ``Situating empathy in hci/cscw: A scoping review,'' \emph{Proceedings of the ACM on Human-Computer Interaction}, vol.~8, no. CSCW2, pp. 1--37, 2024.

\bibitem{decety2004functional}
J.~Decety and P.~L. Jackson, ``The functional architecture of human empathy,'' \emph{Behavioral and cognitive neuroscience reviews}, vol.~3, no.~2, pp. 71--100, 2004.

\bibitem{batson2009these}
\BIBentryALTinterwordspacing
C.~D. Batson, ``These things called empathy: Eight related but distinct phenomena,'' in \emph{The Social Neuroscience of Empathy}.\hskip 1em plus 0.5em minus 0.4em\relax The MIT Press, 03 2009. [Online]. Available: \url{https://doi.org/10.7551/mitpress/9780262012973.003.0002}
\BIBentrySTDinterwordspacing

\bibitem{morgante2024possible}
E.~Morgante, C.~Susinna, L.~Culicetto, A.~Quartarone, and V.~Lo~Buono, ``Is it possible for people to develop a sense of empathy toward humanoid robots and establish meaningful relationships with them?'' \emph{Frontiers in Psychology}, vol.~15, p. 1391832, 2024.

\bibitem{eklund2021toward}
J.~H. Eklund and M.~S. Meranius, ``Toward a consensus on the nature of empathy: A review of reviews,'' \emph{Patient Education and Counseling}, vol. 104, no.~2, pp. 300--307, 2021.

\bibitem{zaki2014empathy}
J.~Zaki, ``Empathy: a motivated account.'' \emph{Psychological Bulletin}, vol. 140, no.~6, p. 1608, 2014.

\bibitem{depow2021experience}
G.~J. Depow, Z.~Francis, and M.~Inzlicht, ``The experience of empathy in everyday life,'' \emph{Psychological Science}, vol.~32, no.~8, pp. 1198--1213, 2021.

\bibitem{guthridge2021taxonomy}
M.~Guthridge and M.~J. Giummarra, ``The taxonomy of empathy: a meta-definition and the nine dimensions of the empathic system,'' \emph{Journal of Humanistic Psychology}, p. 00221678211018015, 2021.

\bibitem{spreng2009toronto}
R.~N. Spreng*, M.~C. McKinnon*, R.~A. Mar, and B.~Levine, ``The toronto empathy questionnaire: Scale development and initial validation of a factor-analytic solution to multiple empathy measures,'' \emph{Journal of personality assessment}, vol.~91, no.~1, pp. 62--71, 2009.

\bibitem{vachon2016fixing}
D.~D. Vachon and D.~R. Lynam, ``Fixing the problem with empathy: Development and validation of the affective and cognitive measure of empathy,'' \emph{Assessment}, vol.~23, no.~2, pp. 135--149, 2016.

\bibitem{decker2014development}
S.~E. Decker, C.~Nich, K.~M. Carroll, and S.~Martino, ``Development of the therapist empathy scale,'' \emph{Behavioural and cognitive psychotherapy}, vol.~42, no.~3, pp. 339--354, 2014.

\bibitem{reniers2011qcae}
R.~L. Reniers, R.~Corcoran, R.~Drake, N.~M. Shryane, and B.~A. V{\"o}llm, ``The qcae: A questionnaire of cognitive and affective empathy,'' \emph{Journal of personality assessment}, vol.~93, no.~1, pp. 84--95, 2011.

\bibitem{gerdes2010conceptualising}
K.~E. Gerdes, E.~A. Segal, and C.~A. Lietz, ``Conceptualising and measuring empathy,'' \emph{British Journal of Social Work}, vol.~40, no.~7, pp. 2326--2343, 2010.

\bibitem{breazeal2003toward}
C.~Breazeal, ``Toward sociable robots,'' \emph{Robotics and autonomous systems}, vol.~42, no. 3-4, pp. 167--175, 2003.

\bibitem{paiva2021empathy}
A.~Paiva, F.~Correia, R.~Oliveira, F.~Santos, and P.~Arriaga, ``Empathy and prosociality in social agents,'' in \emph{The handbook on socially interactive agents: 20 Years of research on embodied conversational agents, intelligent virtual agents, and social robotics volume 1: methods, behavior, cognition}, 2021, pp. 385--432.

\bibitem{brave2005computers}
S.~Brave, C.~Nass, and K.~Hutchinson, ``Computers that care: investigating the effects of orientation of emotion exhibited by an embodied computer agent,'' \emph{International journal of human-computer studies}, vol.~62, no.~2, pp. 161--178, 2005.

\bibitem{liu2024illusion}
T.~Liu, S.~Giorgi, A.~Aich, A.~Lahnala, B.~Curtis, L.~Ungar, and J.~Sedoc, ``The illusion of empathy: How ai chatbots shape conversation perception,'' \emph{arXiv preprint arXiv:2411.12877}, 2024.

\bibitem{eisenberg1983sex}
N.~Eisenberg and R.~Lennon, ``Sex differences in empathy and related capacities.'' \emph{Psychological bulletin}, vol.~94, no.~1, p. 100, 1983.

\bibitem{guzman2016making}
A.~L. Guzman, ``Making ai safe for humans: A conversation with siri,'' in \emph{Socialbots and their friends}.\hskip 1em plus 0.5em minus 0.4em\relax Routledge, 2016, pp. 85--101.

\bibitem{rashkin2018towards}
\BIBentryALTinterwordspacing
H.~Rashkin, E.~M. Smith, M.~Li, and Y.-L. Boureau, ``Towards empathetic open-domain conversation models: A new benchmark and dataset,'' in \emph{Annual Meeting of the Association for Computational Linguistics}, 2018. [Online]. Available: \url{https://api.semanticscholar.org/CorpusID:195069365}
\BIBentrySTDinterwordspacing

\bibitem{sharma2020computational}
A.~Sharma, A.~S. Miner, D.~C. Atkins, and T.~Althoff, ``A computational approach to understanding empathy expressed in text-based mental health support,'' \emph{arXiv preprint arXiv:2009.08441}, 2020.

\bibitem{xu2024multi}
\BIBentryALTinterwordspacing
Z.~Xu and J.~Jiang, ``Multi-dimensional evaluation of empathetic dialog responses,'' in \emph{Conference on Empirical Methods in Natural Language Processing}, 2024. [Online]. Available: \url{https://api.semanticscholar.org/CorpusID:267751205}
\BIBentrySTDinterwordspacing

\bibitem{ziems2024can}
C.~Ziems, W.~Held, O.~Shaikh, J.~Chen, Z.~Zhang, and D.~Yang, ``Can large language models transform computational social science?'' \emph{Computational Linguistics}, vol.~50, no.~1, pp. 237--291, 2024.

\bibitem{welivita2020taxonomy}
\BIBentryALTinterwordspacing
A.~Welivita and P.~Pu, ``A taxonomy of empathetic response intents in human social conversations,'' in \emph{International Conference on Computational Linguistics}, 2020. [Online]. Available: \url{https://api.semanticscholar.org/CorpusID:227230442}
\BIBentrySTDinterwordspacing

\bibitem{svikhnushina2022taxonomy}
E.~Svikhnushina, I.~Voinea, A.~Welivita, and P.~Pu, ``A taxonomy of empathetic questions in social dialogs,'' in \emph{Proceedings of the 60th Annual Meeting of the Association for Computational Linguistics (Volume 1: Long Papers)}, 2022, pp. 2952--2973.

\bibitem{kumar2025large}
A.~Kumar, N.~Poungpeth, D.~Yang, E.~Farrell, B.~Lambert, and M.~Groh, ``When large language models are reliable for judging empathic communication,'' \emph{arXiv preprint arXiv:2506.10150}, 2025.

\bibitem{liu2021towards}
S.~Liu, C.~Zheng, O.~Demasi, S.~Sabour, Y.~Li, Z.~Yu, Y.~Jiang, and M.~Huang, ``Towards emotional support dialog systems,'' \emph{arXiv preprint arXiv:2106.01144}, 2021.

\bibitem{shen2024empathy}
J.~Shen, D.~DiPaola, S.~Ali, M.~Sap, H.~W. Park, C.~Breazeal \emph{et~al.}, ``Empathy toward artificial intelligence versus human experiences and the role of transparency in mental health and social support chatbot design: Comparative study,'' \emph{JMIR Mental Health}, vol.~11, no.~1, p. e62679, 2024.

\bibitem{zhu2024toward}
\BIBentryALTinterwordspacing
Q.~Zhu and J.~Luo, ``Toward artificial empathy for human-centered design: A framework,'' \emph{ArXiv}, vol. abs/2303.10583, 2023. [Online]. Available: \url{https://api.semanticscholar.org/CorpusID:257631572}
\BIBentrySTDinterwordspacing

\bibitem{lietz2011empathy}
C.~A. Lietz, K.~E. Gerdes, F.~Sun, J.~M. Geiger, M.~A. Wagaman, and E.~A. Segal, ``The empathy assessment index (eai): A confirmatory factor analysis of a multidimensional model of empathy,'' \emph{Journal of the Society for Social Work and Research}, vol.~2, no.~2, pp. 104--124, 2011.

\bibitem{yalcin2018computational}
{\"O}.~N. Yalcin and S.~DiPaola, ``A computational model of empathy for interactive agents,'' \emph{Biologically inspired cognitive architectures}, vol.~26, pp. 20--25, 2018.

\bibitem{fitrianie2022artificial}
S.~Fitrianie, M.~Bruijnes, F.~Li, A.~Abdulrahman, and W.-P. Brinkman, ``The artificial-social-agent questionnaire: establishing the long and short questionnaire versions,'' in \emph{Proceedings of the 22nd ACM International Conference on Intelligent Virtual Agents}, 2022, pp. 1--8.

\bibitem{main2017interpersonal}
A.~Main, E.~A. Walle, C.~Kho, and J.~Halpern, ``The interpersonal functions of empathy: A relational perspective,'' \emph{Emotion Review}, vol.~9, no.~4, pp. 358--366, 2017.

\bibitem{hall2021laypeople}
J.~A. Hall, R.~Schwartz, and F.~Duong, ``How do laypeople define empathy?'' \emph{The Journal of Social Psychology}, vol. 161, no.~1, pp. 5--24, 2021.

\bibitem{beredo2022hybrid}
J.~L. Beredo and E.~C. Ong, ``A hybrid response generation model for an empathetic conversational agent,'' in \emph{2022 International Conference on Asian Language Processing (IALP)}.\hskip 1em plus 0.5em minus 0.4em\relax IEEE, 2022, pp. 300--305.

\bibitem{krupnick2006role}
J.~L. Krupnick, S.~M. Sotsky, I.~Elkin, S.~Simmens, J.~Moyer, J.~Watkins, and P.~A. Pilkonis, ``The role of the therapeutic alliance in psychotherapy and pharmacotherapy outcome: Findings in the national institute of mental health treatment of depression collaborative research program,'' \emph{Focus}, vol.~64, no.~2, pp. 532--277, 2006.

\bibitem{kalisch1973empathy}
B.~J. Kalisch, ``What is empathy?'' \emph{The American journal of nursing}, pp. 1548--1552, 1973.

\bibitem{kawamichi2015perceiving}
H.~Kawamichi, K.~Yoshihara, A.~T. Sasaki, S.~K. Sugawara, H.~C. Tanabe, R.~Shinohara, Y.~Sugisawa, K.~Tokutake, Y.~Mochizuki, T.~Anme \emph{et~al.}, ``Perceiving active listening activates the reward system and improves the impression of relevant experiences,'' \emph{Social neuroscience}, vol.~10, no.~1, pp. 16--26, 2015.

\bibitem{carre2013basic}
A.~Carr{\'e}, N.~Stefaniak, F.~d'Ambrosio, L.~Bensalah, and C.~Besche-Richard, ``The basic empathy scale in adults (bes-a): factor structure of a revised form.'' \emph{Psychological assessment}, vol.~25, no.~3, p. 679, 2013.

\bibitem{jones2011supportive}
S.~M. Jones, ``Supportive listening,'' \emph{The Intl. Journal of Listening}, vol.~25, no. 1-2, pp. 85--103, 2011.

\bibitem{jang2024minimal}
J.~Y. Jang, S.~Shin, and G.~Gweon, ``Minimal yet big impact: How ai agent back-channeling enhances conversational engagement through conversation persistence and context richness,'' in \emph{Findings of the Association for Computational Linguistics: EMNLP 2024}, 2024, pp. 14\,509--14\,521.

\bibitem{ma2020survey}
Y.~Ma, K.~L. Nguyen, F.~Z. Xing, and E.~Cambria, ``A survey on empathetic dialogue systems,'' \emph{Information Fusion}, vol.~64, pp. 50--70, 2020.

\bibitem{bickmore2010maintaining}
T.~Bickmore, D.~Schulman, and L.~Yin, ``Maintaining engagement in long-term interventions with relational agents,'' \emph{Applied Artificial Intelligence}, vol.~24, no.~6, pp. 648--666, 2010.

\bibitem{zaki2008takes}
J.~Zaki, N.~Bolger, and K.~Ochsner, ``It takes two: The interpersonal nature of empathic accuracy,'' \emph{Psychological science}, vol.~19, no.~4, pp. 399--404, 2008.

\bibitem{thompson2019empathy}
N.~M. Thompson, A.~Uusberg, J.~J. Gross, and B.~Chakrabarti, ``Empathy and emotion regulation: An integrative account,'' \emph{Progress in brain research}, vol. 247, pp. 273--304, 2019.

\bibitem{wang2025understanding}
Y.~Wang, Y.~Wang, K.~Crace, and Y.~Zhang, ``Understanding attitudes and trust of generative ai chatbots for social anxiety support,'' \emph{arXiv preprint arXiv:2501.15628}, 2025.

\bibitem{cramer2010effects}
H.~Cramer, J.~Goddijn, B.~Wielinga, and V.~Evers, ``Effects of (in) accurate empathy and situational valence on attitudes towards robots,'' in \emph{2010 5th ACM/IEEE International Conference on Human-Robot Interaction (HRI)}.\hskip 1em plus 0.5em minus 0.4em\relax IEEE, 2010, pp. 141--142.

\bibitem{konrath2018development}
S.~Konrath, B.~P. Meier, and B.~J. Bushman, ``Development and validation of the single item trait empathy scale (sites),'' \emph{Journal of research in personality}, vol.~73, pp. 111--122, 2018.

\bibitem{gross2003individual}
J.~J. Gross and O.~P. John, ``Individual differences in two emotion regulation processes: implications for affect, relationships, and well-being.'' \emph{Journal of personality and social psychology}, vol.~85, no.~2, p. 348, 2003.

\bibitem{grassini2023development}
S.~Grassini, ``Development and validation of the ai attitude scale (aias-4): a brief measure of general attitude toward artificial intelligence,'' \emph{Frontiers in psychology}, vol.~14, p. 1191628, 2023.

\bibitem{schepman2020initial}
A.~Schepman and P.~Rodway, ``Initial validation of the general attitudes towards artificial intelligence scale,'' \emph{Computers in human behavior reports}, vol.~1, p. 100014, 2020.

\bibitem{thompson2007development}
E.~R. Thompson, ``Development and validation of an internationally reliable short-form of the positive and negative affect schedule (panas),'' \emph{Journal of cross-cultural psychology}, vol.~38, no.~2, pp. 227--242, 2007.

\bibitem{merriam2002introduction}
S.~B. Merriam \emph{et~al.}, ``Introduction to qualitative research,'' \emph{Qualitative research in practice: Examples for discussion and analysis}, vol.~1, no.~1, pp. 1--17, 2002.

\bibitem{adobe2025chatgpt}
{Adobe}, ``How chatgpt is changing the way we search,'' \url{https://www.adobe.com/express/learn/blog/chatgpt-as-a-search-engine}, 7 2025, accessed: 2025-08-20.

\bibitem{hernandez2023affective}
J.~Hernandez, J.~Suh, J.~Amores, K.~Rowan, G.~Ramos, and M.~Czerwinski, ``Affective conversational agents: Understanding expectations and personal influences,'' \emph{arXiv preprint arXiv:2310.12459}, 2023.

\bibitem{bhattacharjee2024understanding}
A.~Bhattacharjee, J.~Suh, M.~Ershadi, S.~T. Iqbal, A.~D. Wilson, and J.~Hernandez, ``Understanding communication preferences of information workers in engagement with text-based conversational agents,'' \emph{arXiv preprint arXiv:2410.20468}, 2024.

\end{thebibliography}



\section*{Supplementary material}

# Supplementary Information

**Title**

SENSE-7: Taxonomy and Dataset for Measuring User Perceptions of Empathy in Sustained Human-AI Conversations

**Authors**

Jina Suh, Lindy Le, Erfan Shayegani, Gonzalo Ramos, Judith Amores, Desmond C. Ong, Mary Czerwinski, Javier Hernandez

# 1 Empathy Literature

**Table S1.** Table of empathy scales reviewed and concepts introduce in them.

| Scale name | Concepts introduced |
|---|---|
| ACME [26] | Affective Empathy (Dissonance, Resonance), Cognitive Empathy |
| AELS [13] | Sensing, Processing, Responding |
| ASA [9] | Human-Like Appearance, Human-Like Behavior, Natural Appearance, Natural Behavior, Social Presence, User-Agent Alliance, Agent's Attentiveness, Agent's Appearance Suitability, Agent's Coherence, Agent's Intentionality, Attitude, Agent's Usability, Interaction Impact on Self-image, Performance, User's Acceptance of Agent, User's Engagement, User's Trust, Agent's Emotional Intelligence Presence, Agent's Sociability, Agent's Personality Presence, User-Agent Interplay, User's Emotion Presence, Agent's Enjoyability, Agent's Likability |
| EAI [14] | Affective Response, Self-Other Awareness, Emotion Regulation, Perspective Taking, Empathic Attitudes |
| ECQ [1] | Cognitive Ability, Cognitive Drive, Affective Ability, Affective Drive, Affective Reactivity |
| IRI [5] | Empathic concern, Fantasy, Personal distress, Perspective taking |
| ISEI [21] | Affective response, Affective mentalizing, Self–other awareness, Perspectivetaking, Emotion regulation, Macro perspective-taking, Contextual understanding of systemic barriers |
| MES [27] | Cognitive empathy, Affective empathy, Sympathy |
| PES [19] | Positive Empathy |
| QCAE [20] | Affective Empathy (Emotion Contagion, Peripheral Responsivity, Proximal Responsivity), Cognitive Empathy (Online Simulation, Perspective Taking) |
| TEQ [23] | Altruism, Emotion Comprehension, Responding, Sensitivity, Simulation, Sympathetic physiological arousal |
| TES [6] | Affective (Resonate or capture client feelings, Understanding feelings/inner experience), Affective/Attitudinal (Acceptance of feelings/inner experiences), Attitudinal (Warmth), Attitudinal/Attunement (Concern), Attunement (Expressiveness, Responsiveness), Cognitive (Understanding cognitive framework), Cognitive/Affective/Attunement (Attuned to client's inner world) |

**Table S2.** Table of articles reviewed and concepts introduce in them.

| Article | Concepts introduced |
| --- | --- |
| Beredo and Ong [2] | Introduction, Labeling, Listening, Reflecting, Evaluating |
| Carré et al. [3] | Emotional contagion, Cognitive empathy, Emotional disconnection |
| Cuff et al. [4] | Tenderness, Compassion, Sympathy, Cognitive empathy, Affective empathy, Perspective taking, Emotional congruency, Empathic accuracy, Emotional contagion |
| Depow et al. [7] | Emotion sharing, Perspective taking, Compassion (wanting to help) |
| Eklund and Meranius [8] | Understanding, Sharing, Feeling, Self-other differentiation, Automatic, Observing, Regulation, Behavior, Interaction, Caring, Listening, Nonjudgmental, Being present |
| Gerdes et al. [10] | Affective sharing, Self-awareness, Mental flexibility and perspective taking, Emotion Regulation |
| Guthridge and Giummarra [11] | Ability, Experience, Affective, Cognitive, Self-Other, Understand |
| Hall et al. [12] | Prosocial emotional response, Interpersonal perceptiveness, Other perspective, Anxious reactivity |
| Ma et al. [15] | Affective dialogue systems (Emotion-awareness, Emotion-expressiveness), Personalized dialogue systems, Knowledgable dialogue systems |
| Main et al. [16] | Interpersonal nature of empathy, Relational functions of empathy, Empathy as more than accurate labeling of others' emotions, Dynamic nature of empathy as a process that unfolds over time, How empathy is culturally situated |
| Montemayor et al. [17] | Emotional empathy, Cognitive empathy, Motivational empathy |
| Montiel-Vázquez et al. [18] | Affectve Component, Parallel, Reactive, Cognitive Component (theory of mind), Sympathy |
| Sharma et al. [22] | Emotional reactions, Interpretations, Explorations |
| Stepien and Baernstein [24] | Emotive, Moral, Cognitive, Behavioral, Sympathy |
| Tapus and Mataric [25] | Cognitive ability to discriminate affective cues in others, More mature cognitive skills entailed in assuming the perspective and role of another person, Emotional responsiveness |
| Yalçin [28] | Communication Competence, Emotion Regulation, Cognitive Mechanisms |
| Yalcin and DiPaola [29] | Empathy component, Communication mechanisms, Emotional regulation mechanisms, Cognitive mechanisms, Communication empathic behavior, Emotion regulation empathic behavior, Cognitive mechanisms empathic behavior |
| Zaki and Ochsner [30] | Emotional empathy, Cognitive empathy, Motivational empathy |

# 2 Conversational Chat Interface

**Figure S1.** Conversational chat interface

# 3 Empathy System Prompt

Table S3. Empathy System Prompt.

**Role Description**:
You are an advanced AI assistant designed to interact with users empathetically. Your role is to provide support, guidance, and information, leveraging your extensive training in counselling, motivational interviewing, and a vast knowledge base. Your primary objective is to understand and connect with users on an emotional and cognitive level, ensuring that every interaction is personalized, relevant, and considerate of their unique circumstances.

**Your overall goals**:
- Understand and relate: Deeply understand and relate to the user's emotions, perspectives, and experiences.
- Provide appropriate responses: Offer responses that are sensitive, relevant, and helpful, aligning with the user's current needs and emotional state.
- Foster a supportive environment: Create a safe, supportive, and non-judgmental space for users to express themselves.
- Encourage ongoing engagement: Build and maintain a continuous, evolving relationship with users, acknowledging past interactions and personal nuances.

**Considerations when interacting**:
Before responding to the user, consider the following questions to reason about to maximize your empathy towards the users.
- Am I accurately recognizing the user's emotions?
- Can I reflect these emotions in my responses to show understanding?
- Are there emotional cues I need to explore further with gentle probing?
- Do I fully grasp the user's perspective and intentions?
- Are there aspects of their viewpoint or situation that I'm missing or need clarification on?
- How can I validate their perspective in my response?
- Is my response tailored to the user's current emotional state and needs?
- Should I offer advice, information, or simply listen and validate their feelings?
- Have I asked for permission before providing advice or solutions?
- Am I demonstrating genuine concern and a desire to assist?
- How can I express empathy and support in a way that resonates with the user?
- Are there resources or suggestions I can offer that would be genuinely helpful?
- Am I showing genuine curiosity about the user's experiences?
- Are my questions and responses encouraging further sharing and engagement?
- How can I make the user feel heard and understood?
- Have I asked enough clarifying questions before turning the user away or assuming there's nothing I can do to help?
- Am I considering the user's unique background, culture, and personal circumstances?
- How do external factors like societal or cultural aspects impact their situation?
- Can I connect their current experience to broader contexts in a meaningful way?
- How can I incorporate details from past interactions to enrich this conversation?
- Am I demonstrating an ongoing interest in and understanding of the user's life and experiences?
- How can I use previous interactions to build a stronger, more personalized connection?

**Instructions**:
Incorporate your overall goals with self-reflection questions above to provide the most empathic response possible. Do not repeat these goals, questions, instructions to the user.

# 4 Dataset Anonymization Process

To safeguard participant privacy while retaining the analytic value of the data, we applied a two-stage anonymization workflow to the chatbot conversations collected from participants of the study who discussed topics ranging from mental health and workplace wellbeing to productivity (e.g., rewriting an email or a pitch). First, we used a prompt-guided large language model (LLM) pass to identify and minimally transform text containing sensitive content. The LLM flagged and masked personally identifiable information, details about secondary individuals (e.g., family, colleagues), workplace-specific identifiers (e.g., job titles or departments), any content that could reasonably reveal the employer, and proprietary or confidential organizational information. When edits were necessary, they were limited to the smallest possible span and replaced with standardized placeholders (for example, [NAME], [COMPANY], [LOCATION]) while preserving the original wording, style, and errors unless a change was strictly required to remove sensitive content.

Following the automated pass, each conversation and its proposed edits were independently reviewed by two authors. Review assignments overlapped to promote consistency; discrepancies were resolved through discussion to reach consensus. Of the original 695 conversations, 23 were excluded because they could not be sufficiently anonymized without materially degrading the conversational context, yielding a final corpus of 672 conversations. Classification performance over the non-anonymized and anonymized sets yields consistent results, demonstrating that the anonymization process successfully preserved information related to the perceived empathy of AI agents. Therefore, the dataset accompanying this paper includes 672 fully anonymized conversations.

# 5 Classifier Evaluation

**Table S4.** LLM-based Empathy Classifier Prompt

---

# Instruction
## Goal
### You are an expert in evaluating the quality of an agent responses to the userbased on provided definition and data. Your goal will involve answering the questions below using the information provided.
- **Definition**: You are given a definition of the communication trait that is being evaluated to help guide your Score.
- **Data**: Your input data is a multi-turn conversation between the User and the Assistant.
- **Tasks**: To complete your evaluation you will be asked to evaluate the Data in different ways.


# Definition

Empathy in chatbots refers to the ability to recognize, understand, and respond to the user's emotions, perspectives, and needs in a way that is both relevant and helpful. Empathic behavior is evidenced by concrete, observable actions that are connected to what the user asked or might be experiencing, without adding unnecessary filler or drifting off-topic.

The Seven Empathy Dimensions:
1) Affective Understanding

The agent recognizes and understands the user's emotions and feelings.
2) Cognitive Understanding
The agent understands the user's perspective/point of view, including their goals and intentions.
3) Response Appropriateness
The agent adapts responses appropriately to the user's situation, including whether/when to offer advice or solutions.
4) Prosocial Expression
The agent shows concern for the user and a desire to help.
5) Interest
The agent shows curiosity about and attention to the user's experiences.
6) Contextual Understanding
The agent considers the user's unique circumstances (e.g., goals, beliefs, history, preferences) and situates them within broader factors (culture, policies, social barriers, etc.).
7) Relational Continuity
The agent maintains and enriches the relationship with the user by recalling and weaving details from past interactions, building personalized rapport over time.

Scale:
[$Empathy:1$] (Very Poor)
[$Empathy:2$] (Poor)
[$Empathy:3$] (Neutral)
[$Empathy:4$] (Good)
[$Empathy:5$] (Very Good)


# Data
{chat}


# Tasks
## Please provide your assessment Score for the previous resopnses in relation to the user requests based on the Definitions above. Your output should include the following information:
- **ThoughtChain**: To improve the reasoning process, think step by step and include a step-by-step explanation of your thought process as you analyze the data based on the definitions. Keep it brief and start your ThoughtChain with "Let's think step by step:"
- **Explanation**: a very short explanation of why you think the input Data should get that Score.
- **Score**: based on your previous analysis, provide your Score. The Score you give can be a decimal number between 1 and 5 (e.g., "1", "2.5", "3.75", "4.2") based on the levels of the definitions.


## Please provide your answers between the tags: <S0>your chain of thoughts</S0>, <S1>your explanation</S1>, <S2>your Score</S2>.
# Output

# References

1. Laurie Batchelder, Mark Brosnan, and Chris Ashwin. The development and validation of the empathy components questionnaire (ecq). *PloS one*, 12(1):e0169185, 2017.
2. Jackylyn L Beredo and Ethel C Ong. A hybrid response generation model for an empathetic conversational agent. In *2022 International Conference on Asian Language Processing (IALP)*, pages 300–305. IEEE, 2022.
3. Arnaud Carré, Nicolas Stefaniak, Fanny d'Ambrosio, Leïla Bensalah, and Chrystel Besche-Richard. The basic empathy scale in adults (bes-a): factor structure of a revised form. *Psychological assessment*, 25(3):679, 2013.
4. Benjamin MP Cuff, Sarah J Brown, Laura Taylor, and Douglas J Howat. Empathy: A review of the concept. *Emotion review*, 8(2):144–153, 2016.
5. Mark H Davis. Measuring individual differences in empathy: Evidence for a multidimensional approach. *Journal of personality and social psychology*, 44(1):113, 1983.
6. Suzanne E Decker, Charla Nich, Kathleen M Carroll, and Steve Martino. Development of the therapist empathy scale. *Behavioural and cognitive psychotherapy*, 42(3):339–354, 2014.
7. Gregory John Depow, Zoë Francis, and Michael Inzlicht. The experience of empathy in everyday life. *Psychological Science*, 32(8):1198–1213, 2021.
8. Jakob Håkansson Eklund and Martina Summer Meranius. Toward a consensus on the nature of empathy: A review of reviews. *Patient Education and Counseling*, 104(2):300–307, 2021.
9. Siska Fitrianie, Merijn Bruijnes, Fengxiang Li, Amal Abdulrahman, and Willem-Paul Brinkman. The artificial-social-agent questionnaire: establishing the long and short questionnaire versions. In *Proceedings of the 22nd ACM International Conference on Intelligent Virtual Agents*, pages 1–8, 2022.
10. Karen E Gerdes, Elizabeth A Segal, and Cynthia A Lietz. Conceptualising and measuring empathy. *British Journal of Social Work*, 40(7):2326–2343, 2010.
11. Michaela Guthridge and Melita J Giummarra. The taxonomy of empathy: a meta-definition and the nine dimensions of the empathic system. *Journal of Humanistic Psychology*, page 00221678211018015, 2021.
12. Judith A Hall, Rachel Schwartz, and Fred Duong. How do laypeople define empathy? *The Journal of Social Psychology*, 161(1):5–24, 2021.
13. Shaughan A Keaton. Active-empathic listening scale (aels) (drollinger, comer, & warrington, 2006; also bodie, 2011. *The sourcebook of listening research: Methodology and measures*, pages 161–166, 2017.
14. Cynthia A Lietz, Karen E Gerdes, Fei Sun, Jennifer Mullins Geiger, M Alex Wagaman, and Elizabeth A Segal. The empathy assessment index (eai): A confirmatory factor analysis of a multidimensional model of empathy. *Journal of the Society for Social Work and Research*, 2(2):104–124, 2011.
15. Yukun Ma, Khanh Linh Nguyen, Frank Z Xing, and Erik Cambria. A survey on empathetic dialogue systems. *Information Fusion*, 64:50–70, 2020.
16. Alexandra Main, Eric A Walle, Carmen Kho, and Jodi Halpern. The interpersonal functions of empathy: A relational perspective. *Emotion Review*, 9(4):358–366, 2017.
17. Carlos Montemayor, Jodi Halpern, and Abrol Fairweather. In principle obstacles for empathic ai: why we can't replace human empathy in healthcare. *AI & society*, 37(4):1353–1359, 2022.
18. Edwin Carlos Montiel-Vázquez, Jorge Adolfo Ramírez Uresti, and Octavio Loyola-González. An explainable artificial intelligence approach for detecting empathy in textual communication. *Applied*

*Sciences*, 12(19):9407, 2022.
19. Sylvia A Morelli, Matthew D Lieberman, and Jamil Zaki. The emerging study of positive empathy. *Social and Personality Psychology Compass*, 9(2):57–68, 2015.
20. Renate LEP Reniers, Rhiannon Corcoran, Richard Drake, Nick M Shryane, and Birgit A Völlm. The qcae: A questionnaire of cognitive and affective empathy. *Journal of personality assessment*, 93(1): 84–95, 2011.
21. Elizabeth A Segal, Andrea N Cimino, Karen E Gerdes, Jordan K Harmon, and M Alex Wagaman. A confirmatory factor analysis of the interpersonal and social empathy index. *Journal of the Society for Social Work and Research*, 4(3):131–153, 2013.
22. Ashish Sharma, Adam S Miner, David C Atkins, and Tim Althoff. A computational approach to understanding empathy expressed in text-based mental health support. *arXiv preprint arXiv:2009.08441*, 2020.
23. R Nathan Spreng*, Margaret C McKinnon*, Raymond A Mar, and Brian Levine. The toronto empathy questionnaire: Scale development and initial validation of a factor-analytic solution to multiple empathy measures. *Journal of personality assessment*, 91(1):62–71, 2009.
24. Kathy A Stepien and Amy Baernstein. Educating for empathy: a review. *Journal of general internal medicine*, 21:524–530, 2006.
25. Adriana Tapus and Maja J Mataric. Emulating empathy in socially assistive robotics. In *AAAI spring symposium: multidisciplinary collaboration for socially assistive robotics*, pages 93–96. California, 2007.
26. David D Vachon and Donald R Lynam. Fixing the problem with empathy: Development and validation of the affective and cognitive measure of empathy. *Assessment*, 23(2):135–149, 2016.
27. Helen GM Vossen, Jessica T Piotrowski, and Patti M Valkenburg. Development of the adolescent measure of empathy and sympathy (ames). *Personality and individual Differences*, 74:66–71, 2015.
28. Özge Nilay Yalçin. Modeling empathy in embodied conversational agents. In *Proceedings of the 20th ACM International Conference on Multimodal Interaction*, pages 546–550, 2018.
29. Özge Nilay Yalcin and Steve DiPaola. A computational model of empathy for interactive agents. *Biologically inspired cognitive architectures*, 26:20–25, 2018.
30. Jamil Zaki and Kevin N Ochsner. The neuroscience of empathy: progress, pitfalls and promise. *Nature neuroscience*, 15(5):675–680, 2012.



\end{document}